\documentclass[prb,aps,twocolumn,eqsecnum,showpacs]{revtex4}
\usepackage{graphicx,amsmath,bm}

\allowdisplaybreaks

\begin{document}
\title{Spectral functions of two-band spinless fermion and single-band
  spin-1/2 fermion models}

\author{E. Orignac}
\affiliation{Laboratoire de Physique de l'ENS-Lyon, CNRS UMR 5672, 46
  All\'ee d'Italie, 69364 Lyon Cedex 07, France} 
\author{M. Tsuchiizu} 
\author{Y. Suzumura} 
\affiliation{Department of Physics,  Nagoya University, 
 Nagoya 464-8602, Japan}
\begin{abstract}
We examine zero-temperature one-particle spectral functions
  for the one-dimensional two-band spinless fermions 
with different velocities and
general forward-scattering interactions.
By using the bosonization technique
and diagonalizing the model 
to two Tomonaga-Luttinger-liquid 
Hamiltonians, 
we obtain general expressions for the spectral functions
  which are given in terms of the  
 Appell hypergeometric functions.
For the case of identical two-band fermions,
corresponding to the SU(2) symmetric spin-1/2 fermions
 with {repulsive} interactions,
 the spectral functions can be expressed in terms of the 
  Gauss hypergeometric {function} and are shown to  recover the double-peak structure suggesting  the well-known ``spin-charge'' separation.
By tuning the difference in velocities for the two-band fermions, 
we clarify the crossover in spectral functions
   from the ``spin-charge'' separation 
 to the decoupled fermions.
We discuss the relevance of our results to the spin-1/2 
Hubbard model under a magnetic field which can be mapped 
onto two-band  spinless fermions.  
\end{abstract}
\date{\today}
\maketitle

\section{Introduction}

The physical properties of interacting fermions in three dimensions 
can be described using the Landau Fermi liquid concept of fermionic
quasiparticles with renormalized masses and weak effective
interactions.\cite{landau_fermiliquid_theory_static,nozieres_book}
In one dimension, the Landau Fermi
liquid concept is not applicable, and interacting spin-1/2 fermions are
described within the 
 Tomonaga-Luttinger (TL) liquid concept, in which quasiparticles are
 replaced by collective density (or charge for electrons) and spin
 excitations propagating independently with respective velocities
 $u_\rho$ and $u_\sigma$, the so-called spin-charge separation phenomenon.\cite{tomonaga_model,luttinger_model,schulz_houches_revue,voit_bosonization_revue,varma_nonfermiliquid_review} 
Moreover, in the TL liquid,
 the single-fermion excitations are unstable and decay into collective
 modes, leading to 
the sharp suppression  in the
 density of states.\cite{suzumura1980} 
Originally, candidate systems
 for the observation of TL liquid physics have been 
quasi-one-dimensional conductors such as 
the first molecular conductor TTF-TCNQ, \cite{Claessen2002,ito2005}
the Fabre and Bechgaard
 salts,\cite{jerome_organic_review} the blue and purple bronzes, and carbon
 nanotubes.\cite{ijima_nanotubes_synthesis}
 Angle-resolved
 photoemission spectroscopy (ARPES)      
 experiments in quasi-one-dimensional conductors 
 have been used to measure electron spectral functions  and 
probe their TL liquid
 features.\cite{Claessen2002,dardel_photoemission_kmoo3,dardel_photoemission,gweon_bronzes_photoemission,mizokawa2002,ishii03_nanotube_pes,denlinger_arpes_LiMoO,sing2003,gweon03_limo6o17_arpes,grioni2004,vescoli_photoemission_tmtsf}
Self-assembled one-dimensional metallic chains of atoms on
semiconductor surfaces have also been considered as potential
candidates for the observation of TL liquid and spin-charge
separation.\cite{himpsel2001,oncel2008}  
 Initially, the ARPES measurement for gold atom chains on silicon surfaces 
 revealed a double band structure 
and the spin-charge separation was suggested for its origin.
  \cite{Losio2001,Crain2004}
However, recent \textit{spin-resolved} ARPES  experiments 
  on  this system\cite{Okuda2010}  
have shown that the double band nature originates in the 
  spin splitting caused by the Rashba effect. {Besides ARPES,
  magnetotunneling measurements between a wire and a two-dimensional
  electron gas\cite{altland1999,grigera2004} or between two
  wires\cite{peca_tunneling_quantumwire} are also sensitive to the
  TL liquid features of the spectral functions. Experiments on
  quantum wires\cite{auslaender2005,jompol2009} have partially
  confirmed the presence of TL liquid effects in tunneling current
  measurements.}  

More recently, it has been proposed that ultracold atomic gases were
also candidates for the observation of the TL liquid and spin-charge
separation.\cite{recati2003,recati2003a,kollath2006} Indeed, 
atom trapping technology, either optical
\cite{paredes_tonks_optical,kinoshita_tonks_continuous,koehl_nofk} or
magnetic\cite{amerongen2008,bouchoule2009}  has permitted the
  realization of one-dimensional systems of interacting
  particles. In parallel, mixtures of bosonic and fermionic
atoms,\cite{guenter2006,ospelkaus2006} heteronuclear mixtures of
fermionic atoms\cite{wille2008,taglieber2008} as well as 
pseudospin-1/2 fermionic atoms\cite{DeMarco1999} have been
trapped and cooled. Recently, partially polarized 
pseudospin-1/2 fermionic atoms 
have been trapped in an array of 1D tubes\cite{liao2010}. 
Finally, an analog of 
of photoemission spectroscopy for cold atomic gases has been
developed.\cite{stewart2008,jin2009} These developments might permit
the future measurement of spectral functions of one-dimensional trapped
atomic gases and a comparison with theoretical predictions. 

From the theoretical point of view, 
spectral functions $A(q,\omega)$ for the TL liquid at $T=0$  
 have been considered in
Refs.\ \onlinecite{meden_spectral,voit_spectral}. 
Power-law singularities with
exponents depending on the TL-liquid parameters have been predicted at
$\omega=\pm u_\nu |q|$ ($\nu=\rho,\sigma$). 
 At $T>0$, the spectral function, $A(q,\omega)$, is strongly affected 
 by the thermal fluctuation, which  reduces the effect of interaction.
Actually, double peaks due to the spin charge separation 
 move to a single peak with increasing
 temperature.\cite{nakamura_suzumura}
The spectral function of the Hubbard model in one dimension have been
investigated in the limit of $U\to +\infty$
\cite{penc_shadowband,penc1997} with the help of the Ogata-Shiba
wavefunction.\cite{ogata_inf} {More recently, combining a
  bosonization approach for the charge degree of freedom and an exact
  treatment of the spin degrees of freedom has permitted to obtain an
  analytic expression of the spectral function 
{in that}
  limit\cite{matveev2007}.}  
For finite $U$, the Hubbard model has
been considered using exact diagonalizations\cite{favand1997}, quantum
Monte Carlo, \cite{zacher1998,abendschein2006} and
dynamical density-matrix-renormalization-group methods.\cite{benthien2004,bulut2006}   
The spectral functions
of a magnetized TL liquid have been studied
analytically\cite{Miyashita2002,Rabello2002} as well as
numerically.\cite{feiguin2009}  

In the present paper, we calculate the fermion spectral
function in a two-component TL liquid for zero temperature. In
Sec.\ \ref{sec:bosonization}, we recall the bosonization treatment of
the two-band of spinless fermion model,\cite{muttalib1986} 
and derive the expression of the real-space fermion 
Green's function at zero temperature. 
The bosonized Hamiltonian of the two-band
spinless fermion models also describes spin-1/2 fermions in a magnetic
field\cite{frahm_confinv_field,penc_magnetic_field,orso2007} and
mixtures of spinless fermions  with
bosons or fermions\cite{cazalilla03_mixture,mathey2004}. In
Sec.\ \ref{sec:spectral-function}, the spectral functions are obtained.  
  In the case of an SU(2)
invariant model, corresponding to a TL
liquid of spin-1/2 fermions in the absence of magnetic field, 
the fermion spectral function can be expressed in terms of  the Gauss
hypergeometric function {as shown in Ref.~\onlinecite{gogolin_1dbook}} 
 In the non SU(2) invariant case, which might
be achieved experimentally with polarized spin-1/2 neutral
fermions,\cite{liao2010} {or quantum wires under a magnetic field\cite{auslaender2005}}
  the fermion spectral functions can be expressed in
terms of Appell hypergeometric functions. Our approach recovers the previous
results,\cite{meden_spectral,voit_spectral} but also allows to
describe the behavior of the spectral function away from the points
$\omega = \pm u_\nu q$. We discuss the applications
of our results to the Hubbard model under a magnetic field in
Sec.\ \ref{sec:hubbard-model}

\section{Bosonization}
\label{sec:bosonization}

We consider a general two-band model of interacting spinless fermions
defined by: 
\begin{eqnarray}
  H&=&-i \sum_{a=1,2} \int dx v_a (\psi^\dagger_{R,a} \partial_x
  \psi_{R,a} - \psi^\dagger_{L,a} \partial_x
  \psi_{L,a})
\nonumber \\ && {}
 + \sum_a  \int dx g_a \rho_a^2 + g \int dx \rho_1
  \rho_2,    
  \label{eq:hamiltonian}
\end{eqnarray}
where $\psi_{R,a}$ and $\psi_{L,a}$ respectively annihilate one
right moving and left moving fermion in band $a$, $\rho_a =  \psi^\dagger_{R,a} 
  \psi_{R,a} + \psi^\dagger_{L,a}   \psi_{L,a}$, $v_a$ is the velocity
  of fermions in band $a \in \{1,2\}$  and $g_a$ the strength of
  interband interaction. The model includes only the forward
  scattering interaction and
  not the backward scattering 
  $\sim \psi_{R,1}^\dagger \psi_{L,1}^{} \psi^\dagger_{L,2} \psi_{R,2}^{}$. 
This assumption is justified when the two bands have different Fermi
wavevectors or when the backward scattering interactions are
irrelevant. 

The model (\ref{eq:hamiltonian}) can be bosonized perturbatively, yielding:
\begin{eqnarray}
  H&=&\sum_{a=1,2} \int \frac{dx}{2\pi} \left[u_a K_a (\pi \Pi_a)^2 + \frac
    {u_a}{K_a} (\partial_x \phi_a)^2 \right] 
\nonumber \\ && {}
+ \frac g {\pi^2} \int dx
  \partial_x \phi_1 \partial_x \phi_2 ,
  \label{eq:bosonized-perturb}
\end{eqnarray}
with $[\phi_a(x),\Pi_b(x')]=i \delta_{a,b} \delta(x-x')$ and $a,b \in
\{1,2\}$. In Eq.\ (\ref{eq:bosonized-perturb}), we have:
\begin{subequations}
\begin{eqnarray}
  u_a^2&=&v_a\left(v_a+\frac{2g_a}{\pi}\right), \\
  K_a&=&\left(1+\frac{2g_a}{\pi  v_a }\right)^{-1/2} .
\end{eqnarray}%
\label{eq:param-perturb}%
\end{subequations}
The intraband interaction is included in $K_1$ and $K_2$.
The case $K_a < 1$ ($K_a > 1$) 
corresponds to the repulsive (attractive) interaction.

More general models (for instance lattice models) can also be considered with
bosonization.
A non-perturbative formulation leads to a Hamiltonian\cite{orignac2010_mix} 
\begin{eqnarray}
  \label{eq:bosonized}
  H=\sum_{a,b} \int \frac{dx}{2\pi} \left[ \pi^2 M_{ab} \Pi_a \Pi_b +
    N_{ab} \partial_x \phi_a \partial_x \phi_b \right],   
\end{eqnarray}
where the matrices $M$ and $N$  are
real symmetric and are defined in terms of the variations 
of the ground state energy $E_{GS}$ of a finite system of size $L$  
from (respectively)  change  of boundary
conditions $\psi_a(L)=e^{i\varphi_a} \psi_a(0)$  and change of
particle densities $\rho_a=N_a/L$:       
\begin{eqnarray}
 \label{eq:definition-M}
    M_{ab} = \pi L \frac{\partial^2
    E_{GS}}{\partial\varphi_a \partial\varphi_b}, \\
  \label{eq:definition-N}
  N_{ab} = \frac 1 {\pi L}  \frac{\partial^2
    E_{GS}}{\partial\rho_a \partial\rho_b}.  
\end{eqnarray} 

The spectrum of the general bosonized Hamiltonian~(\ref{eq:bosonized})
is obtained by a 
linear transformation of the fields $\Pi_a$ and $\phi_a$: 
\begin{eqnarray}
  \label{eq:linear-transform}
  \Pi_b &=& \sum_{\beta}  P_{b\beta} \tilde{\Pi}_\beta, \\ 
  \phi_a &=& \sum_{\alpha} Q_{a\alpha} \tilde{\phi}_\alpha, 
\end{eqnarray}
where  $P{}^t Q=1$ in order to preserve the canonical commutation
relations.{\cite{hikihara2005}}  The matrices $P$ and $Q$ are calculated explicitly by applying a
succession of linear transformations.  First, the 
matrix $M$ is diagonalized by a rotation $R_1$ ($\Delta_1={}^t R_1 M
R_1$)  which transforms the
matrix $N$ into 
$N_1={}^t R_1 N R_1$. 
Hereafter we denote $\bm{\phi}={}^t(\phi_1,\phi_2)$ and 
$\bm{\Pi}={}^t(\Pi_1,\Pi_2)$.
By the transformation $\bm{\Pi}=R_1
\bm{\Pi}_1$ and $\bm{\phi} = R_1 \bm{\phi}_1$, the Hamiltonian is thus
transformed into:
\begin{eqnarray}
  H=\int \frac{dx}{2\pi} \left[ \pi^2 {}^t \bm{\Pi}_1 \Delta_1 \bm{\Pi}_1 + {}^t
    (\partial_x \bm{\phi}_1) N_1 (\partial_x \bm{\phi}_1) \right].
\end{eqnarray}
Using a second transformation $\bm{\Pi}_1= \Delta_1^{-1/2} \bm{\Pi}_2$ and 
$\bm{\phi}_1=\Delta_1^{1/2} \bm{\phi}_2$,
 the Hamiltonian becomes: 
 \begin{equation}
  H=\int \frac{dx}{2\pi} \left[ \pi^2 \, {}^t \bm{\Pi}_2  \bm{\Pi}_2 + {}^t
    (\partial_x \bm{\phi}_2) 
\Delta_1^{1/2} N_1 \Delta_1^{1/2}
    (\partial_x \bm{\phi}_2) \right].  
\end{equation}
As the matrix $\Delta_1^{1/2} N_1 \Delta_1^{1/2}$ is symmetric, it can be
diagonalized by a rotation $R_2$ i.e., $\Delta_1^{1/2} N_1
\Delta_1^{1/2}= R_2 \Delta_2 {}^t R_2$. 
Writing $\bm{\Pi}_2=R_2 \bm{\Pi}_3$ and
$\bm{\phi}_2=R_2 \bm{\phi}_3$, we find that the  Hamiltonian 
can be diagonalized as
\begin{eqnarray}
  \label{eq:diago-ham}
H=  \int \frac{dx}{2\pi} \left[ \pi^2 \, {}^t \bm{\Pi}_3  \bm{\Pi}_3 + {}^t
    (\partial_x \bm{\phi}_3) \Delta_2 (\partial_x \bm{\phi}_3) \right],  
\end{eqnarray}
 in which the modes are decoupled. Finally, we can rescale the fields
 $\bm{\Pi}_3 = (\Delta_2)^{1/4} \tilde{\bm{\Pi}}$ and 
$\bm{\phi}_3 = (\Delta_2)^{-1/4} \tilde{\bm{\phi}}$ to write: 
 \begin{equation}\label{eq:bosonized-diagonal}
   H=  \int \frac{dx}{2\pi}
 \left[ \pi^2 \, {}^t \tilde{\bm\Pi}  (\Delta_2)^{1/2}
  \tilde{\bm\Pi}
  + {}^t (\partial_x \tilde{\bm\phi})  (\Delta_2)^{1/2} 
  (\partial_x \tilde{\bm\phi}) \right].
 \end{equation}
In this last equation, the elements on the diagonal of
$(\Delta_2)^{1/2}$ are the velocities $u_\beta$ of the decoupled modes of the
Hamiltonian~(\ref{eq:bosonized}). 
The stability of the multicomponent TL liquid state requires
that the velocities in (\ref{eq:diago-ham}) are real, 
i.e., that the matrix $M N$ has only positive eigenvalues. 
The transformations can be written explicitly as:
\begin{eqnarray}\label{eq:transformation} 
  P&=&R_1   \Delta_1^{-1/2} R_2 (\Delta_2)^{1/4}, \\
  Q &=& R_1   \Delta_1^{1/2} R_2(\Delta_2)^{-1/4},  
\end{eqnarray}
and we have: ${}^t P M P =(\Delta_2)^{1/2}$ and ${}^t Q N Q =
(\Delta_2)^{1/2}$. 
This implies in particular that: ${}^t P MN Q = \Delta_2$ i.e. $Q^{-1} MN
Q=\Delta_2$, and by taking the transpose, $P^{-1} NM P=\Delta_2$. 
The excitation velocities $u_\pm$ where
$\Delta_2=\mathrm{diag}(u_+^2,u_-^2)$ are obtained as
\begin{widetext}
\begin{eqnarray}
u_\pm^2
&=&
\frac{u_1^2 + u_2^2}{2} + M_{12}N_{12}
\pm \sqrt{
\left(
\frac{u_1^2 - u_2^2}{2}
\right)^2
+\left(M_{11}N_{12} + M_{12}N_{22}\right)
 \left(M_{21}N_{11} + M_{22}N_{21}\right)
},
\end{eqnarray}
\end{widetext}
where 
$u_1^2 \equiv N_{11} M_{11}$ and 
$u_2^2 \equiv N_{22} M_{22}$.

The diagonalization method allows us to derive also expressions for
the Green's functions of the chiral fields,
\begin{eqnarray}
  \label{eq:chiral}
\left\{
\begin{array}{l}
  \phi_{R,a}=\phi_a-\theta_a \\
  \phi_{L,a}=\phi_a + \theta_a
\end{array}
\right.
, \quad
\left\{
\begin{array}{l}
  \tilde\phi_{R,a}=\tilde\phi_a - \tilde\theta_a \\
  \tilde\phi_{L,a}=\tilde\phi_a + \tilde\theta_a
\end{array}
\right.
,
\end{eqnarray}
where $\theta_a=\pi \int^x dx' \Pi_a(x')$.
The field operators are expressed as
\begin{subequations}
\begin{eqnarray}
\psi_{R,a}(x,t)&=& \frac{1}{\sqrt{2\pi\alpha}} e^{-i\phi_{R,a}(x,t)}, 
\\
\psi_{L,a}(x,t)&=& \frac{1}{\sqrt{2\pi\alpha}} e^{i\phi_{L,a}(x,t)},
\end{eqnarray}
\end{subequations}
where $\alpha$ is the short-distance cutoff.
 The Hamiltonian (\ref{eq:bosonized-diagonal}) 
can be reexpressed in terms of non-interacting chiral
 fields: 
 \begin{equation}
   H=\int \frac{dx}{4\pi} \left[{}^t (\partial_x \tilde{\bm\phi}_R)
     \Delta_2^{1/2}  (\partial_x \tilde{\bm\phi}_R) 
   + {}^t (\partial_x \tilde{\bm\phi}_L)
     \Delta_2^{1/2}  (\partial_x \tilde{\bm\phi}_L) \right], 
 \end{equation}
with the transformation:
\begin{eqnarray}
  \bm\phi_R&=&\frac 1 2 \left[ (Q + P)
    \tilde{\bm\phi}_R + (Q  - P) 
    \tilde{\bm\phi}_L \right], \\ 
   \bm\phi_L&=&\frac 1 2 \left[ (Q - P)
    \tilde{\bm\phi}_R + (Q  + P) 
    \tilde{\bm\phi}_L \right],     
\end{eqnarray}
where $\tilde{\bm\phi}_R={}^t(\phi_{R,1},\phi_{R,2})$ and 
$\tilde{\bm\phi}_L={}^t(\phi_{L,1},\phi_{L,2})$.

For the model (\ref{eq:hamiltonian}), 
the off-diagonal term of $N$ is given by
$N_{12}=g/\pi$ and 
there is no interband current-current
interaction, i.e., $M_{12}=0$.
In this case, 
the explicit forms of the matrices $P$ and $Q$ for the 
model (\ref{eq:hamiltonian}) can be 
expressed in a compact form.
Since $M_{12}=0$,  the matrix $R_1$ becomes unit
matrix, and $\Delta_1=M=\mathrm{diag}(u_1K_1, u_2K_2)$.
The matrix $\Delta_2$ is given by
$\Delta_2=\mathrm{diag}(u_+^2,u_-^2)$ where
\begin{equation}
u_{\pm}^2
=
\frac{u_1^2+u_2^2}{2}
\pm
\sqrt{
\left(\frac{u_1^2-u_2^2}{2}\right)^2
+\left(\frac{g}{\pi}\right)^2 u_1 K_1 u_2 K_2
}.
\label{eq:velocity}
\end{equation}  
We note that 
 the velocities $u_{\pm}$  depend on $g^2$, i.e., 
do not depend on the sign of $g$. 
From Eq.\ (\ref{eq:transformation}), we obtain
\begin{eqnarray}
  P&=&
\left(
\begin{array}{cc}
\sqrt{\frac{u_{+}}{u_1K_1}}\cos \frac{\alpha}{2} &
 -\sqrt{\frac{u_{-}}{u_1K_1}} \sin \frac{\alpha}{2}
\\ 
\sqrt{\frac{u_{+}}{u_2K_2}} \sin \frac{\alpha}{2} & 
\sqrt{\frac{u_{-}}{u_2K_2}} \cos \frac{\alpha}{2} 
\end{array}
\right)
, \\
  Q &=& 
\left(
\begin{array}{cc}
\sqrt{\frac{u_1K_1}{u_+}} \cos \frac{\alpha}{2} & 
-\sqrt{\frac{u_1K_1}{u_-}} \sin \frac{\alpha}{2}
\\ 
\sqrt{\frac{u_2K_2}{u_+}} \sin \frac{\alpha}{2} & 
\sqrt{\frac{u_2K_2}{u_-}} \cos \frac{\alpha}{2} 
\end{array}
\right),
\end{eqnarray}
where 
$\tan \alpha = 2(g/\pi)
\sqrt{ u_1 K_1 u_2 K_2}/(u_1^2-u_2^2)$.
From Eq.\ (\ref{eq:velocity}), 
the stability condition is given by
\begin{eqnarray}
\label{eq:cond_uc}
u_1 u_2 >  \left(\frac{g}{\pi}\right)^2K_1K_2.
\end{eqnarray}
In the following analysis, we restrict ourselves to the case
  of $u_1>u_2$, and Eq.\ (\ref{eq:cond_uc}) gives the lower condition
  for $u_2$, i.e.,
  $u_2 > u_{2c}\equiv (g/\pi)^2 K_1 K_2 / u_1$.

\section{Phase diagram}

We derive a phase diagram for the model
(\ref{eq:hamiltonian}),
 by examining asymptotic behavior of correlation functions:
\begin{eqnarray}
 \left< O_A(x) O_A(0) \right> \sim x^{-\eta_A},
\end{eqnarray}
where the $O_A$'s represent the order parameters and 
the $\eta_A$'s  the corresponding exponents.
We restrict ourselves to the simplest case $g_1=g_2$ in 
 Eq.\ (\ref{eq:hamiltonian}), and assume $v_2/v_1 \le 1$ without 
loss of generality. 
In the TL-liquid state, 
the state with the smallest exponent represents the (quasi)
long-range-ordered state 
and thus we can determine the phase diagram.
As possible order parameters, 
we can consider the intraband density wave (DW) and 
pairing of superconducting state (SC), which are given by 
 $O_{\mathrm{DW}a} \propto \exp( i 2 \phi_a)$ and 
 $O_{\mathrm{SC}a} \propto \exp( i 2 \theta_a)$.
Their respective exponents $\eta_A$
 are given by

\begin{subequations}
\label{eq:eta}
\begin{eqnarray}
\eta_\mathrm{DW1} &=& 
\frac{u_1K_1}{u_+ u_-}
\left(
 u_+ + u_- - \frac{u_1^2-u_2^2}{u_+ +u_-}
\right),
\\
\eta_\mathrm{DW2} &=& 
\frac{u_2K_2}{u_+ u_-}
\left(
 u_+ + u_- + \frac{u_1^2-u_2^2}{u_++u_-}
\right),
\\
\eta_\mathrm{SC1} &=& 
\frac{1}{u_1K_1}
\left(
 u_+ + u_- + \frac{u_1^2-u_2^2}{u_+ +u_-}
\right),
\\
\eta_\mathrm{SC2} &=& 
\frac{1}{u_2K_2}
\left(
 u_+ + u_- - \frac{u_1^2-u_2^2}{u_++u_-}
\right).
\end{eqnarray}  
\end{subequations}

In addition, we can also consider 
the order parameters for the interband DW and interband SC states, given by
  $O_\mathrm{interband \, DW} = 
\psi_{R,1}^\dagger \psi_{L,2}
\propto e^{i\phi_{R,1}+i\phi_{L,2}}$ 
and
$O_\mathrm{interband \, SC} = 
\psi_{R,1} \psi_{L,2}
\propto 
e^{-i\phi_{R,1}+i\phi_{L,2}}$.
The corresponding exponents are given by
\begin{subequations}
\label{eq:eta-interband}
\begin{eqnarray}
\eta_{\mathrm{interband \, DW}} &=& 
 \frac{\eta_\mathrm{DW1}}{4} 
+ \frac{\eta_\mathrm{DW2}}{4}
+ \frac{\eta_\mathrm{SC1}}{4} 
+ \frac{\eta_\mathrm{SC2}}{4}
\nonumber \\ && {}
-\frac{(g/\pi)}{u_+ +u_-}
\left( 1 + \frac{u_1K_1 u_2K_2}{u_+ u_-} \right), \qquad
\\
\eta_{\mathrm{interband \, SC}} &=& 
 \frac{\eta_\mathrm{DW1}}{4} 
+ \frac{\eta_\mathrm{DW2}}{4}
+ \frac{\eta_\mathrm{SC1}}{4} 
+ \frac{\eta_\mathrm{SC2}}{4}
\nonumber \\ && {}
+\frac{(g/\pi)}{u_+ +u_-}
\left( 1 + \frac{u_1K_1 u_2K_2}{u_+ u_-} \right) . \qquad
\end{eqnarray}  
\end{subequations}

\begin{figure}[t]
\includegraphics[width=7cm]{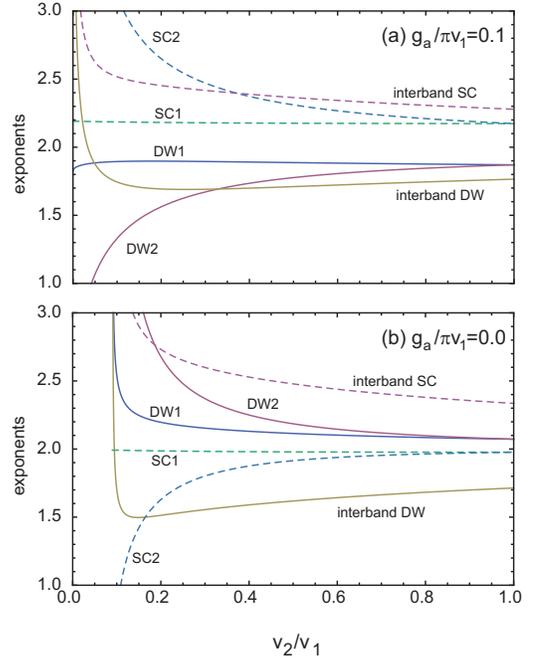}
\caption{
(Color online)
Exponents $\eta_A$ for possible order parameters
as a function of $v_2/v_1$ for $g_a/(\pi v_1) = 0.1$ (a) 
  and  $g_a/(\pi v_1) = 0.0$ (b) with fixed $g/(\pi v_1)=0.3$.
}
\label{fig:exponent-v1v2}
\end{figure}

\begin{figure}[t]
\includegraphics[width=7cm]{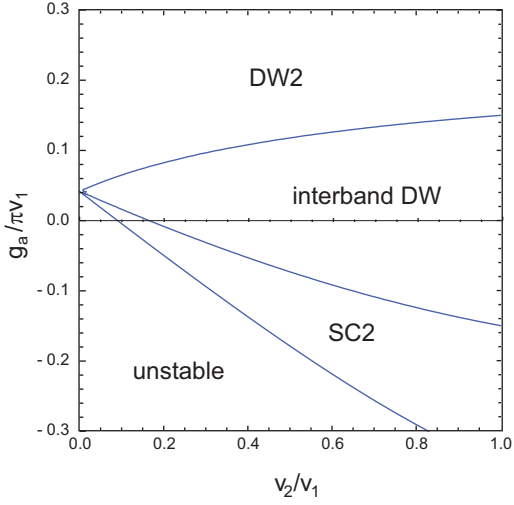}
\caption{
(Color online)
Phase diagram on a plane of $v_2/v_1$ and $g_a/(\pi v_1)$ 
with fixed $g/(\pi v_1)=0.3$.
In the ``unstable'' region, 
  $u_2$ becomes imaginary.
}
\label{fig:phasediagram-v1v2}
\end{figure}

In Fig.~\ref{fig:exponent-v1v2},  
these exponents are shown 
as a function of $v_2/v_1$, for repulsive interactions.
For  $v_2/v_1 =1$, 
 the interband DW state becomes dominant if  $g>2g_a$ [see Fig.~\ref{fig:exponent-v1v2} (a)] 
  while the DW2 state becomes dominant if $g<2g_a$.
For decreasing  $v_2$, 
  the interband DW state is 
  unfavorable and
 instead the DW2  state becomes dominant since the effect of the
  intraband interaction $g_a/v_2$ is enhanced.
On the other hand, 
  if the intraband interactions are absent ($g_a=0$) [see Fig.~\ref{fig:exponent-v1v2}
  (b)],  
the  DW2 state is no longer enhanced for small $v_2$ and instead 
 the exponent $\eta_\mathrm{SC2}$ decreases with decreasing $v_2$
  and the  SC2 state becomes most dominant state 
  and finally the two-component TLL state becomes unstable 
  ($u_-^2<0$).
In Fig.\ \ref{fig:phasediagram-v1v2}, 
the phase diagram on the plane of $v_2/v_1$ and $g_a/(\pi v_1)$ 
is shown  with fixed $g/(\pi v_1)=0.3$.
In the analogy to the spinful electron model, 
  the DW2, interband DW, and SC2 states corresponds to the conventional
  charge-density-wave, spin-density-wave, and triplet SC states,
  respectively.
The region of the interband DW state 
becomes narrow with decreasing $v_2$ and 
shrinks at the point 
$(v_2/v_1,g_a/(\pi v_1))\approx (0, 0.04)$,
  where the SC2, DW2 and interband DW states have the same exponent.

We note that for $g=0$, we can recover the behavior 
that the DW2 state becomes dominant for $g_a>0$ while 
  the SC2 state becomes dominant for $-\pi v_2/2 < g_a<0$,
  and the system becomes unstable 
  if $g_a<-\pi v_2/2$. 
We note that the expressions 
 (\ref{eq:param-perturb}) are valid only for weak interactions, 
 and thus the precise 
  determination of  phase diagram for the region  $g_a/v_2 \gg 1$
  in Fig.\ \ref{fig:phasediagram-v1v2} is beyond the present approach.
From the qualitative considerations,
the {following} modifications to the phase diagram can be expected.
Since the parameter $K_2$ would take a nonzero value  for $g_a/v_2\to \infty$,  
the condition (\ref{eq:cond_uc}) cannot be fulfilled 
  for small $v_2/v_1$, and then
  the unstable region always appears in the small limit of  $v_2/v_1$ 
  for all values of $g_a$.
Furthermore, by noting that 
the unstable region is adjacent to the SC2 state, 
it is expected that
the SC2 state is obtained even for large $g_a$ region and is
 located in between the DW2 state and the unstable region.

\section{Spectral function}
\label{sec:spectral-function}
The spectral function is obtained from
\begin{eqnarray}
  \label{eq:spectral-def}
  A_a(k,\omega)=
-\frac{1}{\pi}
\mathrm{Im} \int dx dt e^{-i(kx-\omega t)}
  G_{a}(x,t),
\end{eqnarray}
where $G_a(x,t)$ is the 
 retarded fermion Green's
function 
$G_{a}(x,t)=-i \theta(t) \langle \{ \psi_{a}(x,t), \psi^\dagger_{a}(0,0)\} \rangle,  $.
Introducing the right moving and left moving components,
$\psi_a(x,t)=e^{ik_{F,a} x} \psi_{R,a}(x,t)+e^{-ik_{F,a} x}
\psi_{L,a}(x,t)$, the chiral Green's functions ($\nu=R,L$) are given by
\begin{eqnarray}
  G_{\nu,a}(x,t)&=&-i \theta(t) [F_{\nu,a}(x,t)+F_{\nu,a}(-x,-t)], 
\end{eqnarray}
where 
$F_{\nu,a}(x,t)= \langle \psi_{\nu,a}(x,t)  \psi^\dagger_{\nu,a}(0,0) \rangle$.
Then the spectral function can be decomposed into the 
four contributions:
\begin{eqnarray}
  A_a(k,\omega)&=&
A_{R,a}(k-k_{F,a},\omega)+A_{L,a}(k+k_{F,a},\omega), 
\nonumber \\ 
 A_{\nu,a}(q,\omega)&=& 
I_{\nu,a}(q,\omega)+I_{\nu,a}(-q,-\omega),
\label{eq:AtoI}
\end{eqnarray}
where
\begin{equation}
\label{eq:spectral-decomp} 
I_{\nu,a}(q,\omega)=
\frac{1}{2\pi}
\int_{-\infty}^\infty dt e^{i\omega t}
\int_{-\infty}^\infty dx e^{- i qx } F_{\nu,a}(x,t) .
\end{equation}

The direct calculation of the Green's functions for the phase variables yields
\begin{widetext}
\begin{eqnarray}
&& \langle(\theta_a(x,t)-\phi_a(x,t))(\theta_a(0,0)-\phi_a(0,0))\rangle_{conn.}
=\sum_{\beta=\pm} \nu_{a,\beta}  
    \ln\left[\frac{\alpha}{\alpha+ i (u_\beta t
    -x)}\right]
  +  \nu'_{a,\beta} \ln\left[\frac{\alpha}{\alpha+ i (u_\beta t
			+x)}\right], 
\\
&& \langle(\theta_a(x,t)+\phi_a(x,t))(\theta_a(0,0)+\phi_a(0,0))\rangle_{conn.}
=\sum_{\beta=\pm} \nu'_{a,\beta}  \ln\left[\frac{\alpha}{\alpha+ i (u_\beta t
    -x)}\right] +   \nu_{a,\beta} \ln\left[\frac{\alpha}{\alpha+ i (u_\beta t
    +x)}\right],
\end{eqnarray} 
where the exponents $\nu$s are given by
\begin{eqnarray}
\nu_{a,\beta}= \frac{1}{4} (P_{a\beta}+Q_{a\beta})^2 , \quad
\nu'_{a,\beta}= \frac{1}{4}(P_{a\beta}-Q_{a\beta})^2 .
\label{eq:exponent}
\end{eqnarray} 
Thus the one-particle Green's functions are expressed as
\begin{subequations}
  \label{eq:fermion-correlators}
\begin{eqnarray}
  \langle \psi_{R,a}(x,t) \psi^\dagger_{R,a}(0,0) \rangle &=& \frac 1
  {2\pi \alpha} \prod_\beta  \left[\frac{\alpha}{\alpha+ i (u_\beta t
    -x)}\right]^{\nu_{a,\beta}} \left[\frac{\alpha}{\alpha+ i (u_\beta t
    +x)}\right]^{\nu'_{a,\beta}}, 
\\ 
   \langle \psi_{L,a}(x,t) \psi^\dagger_{L,a}(0,0) \rangle 
&=& \frac 1
  {2\pi \alpha} \prod_\beta  \left[\frac{\alpha}{\alpha+ i (u_\beta t
    -x)}\right]^{\nu'_{a,\beta}} \left[\frac{\alpha}{\alpha+ i (u_\beta t
    +x)}\right]^{\nu_{a,\beta}}.  
\end{eqnarray}
\end{subequations}
Substituting the expression from Eq.\ (\ref{eq:fermion-correlators}) into
Eq.\ (\ref{eq:spectral-decomp}), we obtain the integral form for the
spectral function. 
We note that we have the identity: $\sum_\beta
(\nu_{a,\beta}-\nu'_{a,\beta})=1$. 
The spectral function is obtained from the integral:  
 \begin{equation}\label{eq:integral} 
    I_{R,a}(q,\omega)=
\frac{\alpha^{\bar \nu_{a}-1}}{(2\pi)^2}
     \int  dx \, dt
\frac{ e^{i(\omega t - q x)} }   
     {[\alpha + i(u_+ t -x)]^{\nu_{a,+}} 
       [\alpha +
      i(u_- t -x)]^{\nu_{a,-}} [\alpha +
      i(u_+ t +x)]^{\nu'_{a,+}}  [\alpha +
      i(u_- t +x)]^{\nu'_{a,-}}},      
  \end{equation} 
\end{widetext}
where $\bar \nu_a \equiv (\nu_{a,+}+\nu_{a,-}+\nu_{a,+}'+\nu_{a,-}')$.
A similar expression for left moving fermions with $\nu_{a,\pm}$
and $\nu'_{a,\pm}$ interchanged.

\subsection{SU(2) symmetric model}
\label{sec:su2}

Let us first consider the case with $SU(2)$ symmetry and repulsive
interactions, {and show how the expressions in terms of Gauss
  hypergeometric functions are recovered.\cite{gogolin_1dbook}}  The bosonized
Hamiltonian reads 
$H=H_\rho+H_\sigma$ where  
\begin{eqnarray}
  H_\rho&=&\int \frac{dx}{2\pi} \left[ u_\rho K_\rho (\pi \Pi_\rho)^2 +
    \frac  {u_\rho} {K_\rho} (\partial_x \phi_\rho)^2 \right], \\
  H_\sigma&=&\int \frac{dx}{2\pi} \left[u_\sigma K_\sigma (\pi \Pi_\sigma)^2 +
    \frac  {u_\sigma} {K_\sigma} (\partial_x \phi_\sigma)^2 \right] 
\nonumber \\ && {}
+
  \frac{2g_{1\perp}}{(2\pi\alpha)^2} \int dx \cos \sqrt{8}\phi_\sigma,  
\end{eqnarray}
with $u_\rho > u_\sigma$, $K_\rho <1$ and $K_\sigma>1$ for repulsive
interactions. Under the
renormalization group, $K_\sigma,g_{1\perp}$ flow to fixed point
values $K_\sigma^*=1, g_{1\perp}^*=0$.    The resulting fixed point
Hamiltonian is therefore in the diagonalized form of Eq.\ 
(\ref{eq:bosonized-diagonal}). We will approximate the spectral
function of this model by replacing the exact Green's function by its
fixed point value. This amounts to neglect logarithmic corrections. 
Within this approximation, the spectral function  of right-moving
fermions
 of spin $s=\uparrow,\downarrow$
 is given by
$A_{R,s}(q,\omega)=I_{R}(q,\omega)+I_R(-q,-\omega)$ 
with
\begin{eqnarray}
&& I_{R}(q,\omega)=
\frac{\alpha^{2\gamma_\rho}}{(2\pi)^2}
\int dx dt\, e^{i(\omega t -q x)}
\nonumber \\ && {}\qquad \times
\frac{1}{[\alpha+ i(u_\rho t -x)]^{\gamma_\rho +1/2}  }
\frac{1}{[\alpha+ i(u_\sigma t -x)]^{1/2}}
\nonumber \\ && {}\qquad \times
\frac{1}{[\alpha+ i(u_\rho t +x)]^{\gamma_\rho}}, 
  \label{eq:su2-spectral}
\end{eqnarray}
where $\gamma_\rho=(K_\rho+K_\rho^{-1}-2)/8$. \footnote{Actually, we could
  consider the more general case of a $SU(N)$ symmetric model with
  $N>2$. The model still has spin-charge separation, but the spin part
  of the correlation decays with an exponent $(N-1)/N$ instead of
  $1/2$ while the charge part has exponent {$(2 \gamma_\rho+1)/N$} for the
  right moving factor, {$2 \gamma_\rho/N$} for the right moving factor. 
The computation of the spectral function proceeds in exactly the 
same manner as in the $SU(2)$ case, albeit with the replacement
{$v^{\gamma_\rho-1/2}(1-v)^{-1/2}/\Gamma(\gamma_\rho+1/2)\Gamma(1/2)
\to
v^{(2\gamma_\rho+1)/N-1}(1-v)^{-1/N}/\Gamma((2\gamma_\rho+1)/N)\Gamma(1-1/N)$} 
in Eq.~(\ref{eq:su2-master}), leading again an expression in terms of
Gauss hypergeometric functions.}      
The correspondence between Eqs.\ (\ref{eq:integral}) and
  (\ref{eq:su2-spectral}) is given by 
$u_+ = u_\rho$, $u_-=u_\sigma$, $\nu_{a,+}=(\gamma_\rho+1/2)$, 
  $\nu_{a,-}=1/2$, $\nu_{a,+}'=\gamma_\rho$, and $\nu_{a,-}'=0$.
To calculate the integral in Eq.~(\ref{eq:su2-spectral}), we use the Feynman
representation\cite{lebellac_qft}:
 \begin{equation}\label{eq:feynman} 
    \frac{1}{A_1^{\nu_1} A_2^{\nu_2}}
    =\frac{\Gamma(\nu_1+\nu_2)}{\Gamma(\nu_1) \Gamma(\nu_2)} \int_0^1
    dw \frac{w^{\nu_1-1} (1-w)^{\nu_2-1}}{\left[A_1 w +
        A_2(1-w)\right]^{\nu_1+\nu_2}},   
  \end{equation} 
which is valid for $\nu_1,\nu_2>0$. Then, with the help of
(\ref{eq:transformed}),  we obtain:
\begin{widetext}
\begin{equation}
 I_{R}(q,\omega)=\frac{\Gamma(\gamma_\rho+1)}{4\pi^2 \Gamma(\gamma_\rho+1/2)\Gamma(1/2)}
 \int dx dt e^{i(\omega t -q x)}\int_0^1 dv \frac{v^{\gamma_\rho-1/2}
   (1-v)^{-1/2} \alpha^{\gamma_\rho}}{\{\alpha+i[v u_\rho + (1-v)
   u_\sigma] t - ix \}^{\gamma_\rho + 1}} \left[\frac{\alpha}{\alpha+
       i(u_\rho t +x)}\right]^{\gamma_\rho}.  
\end{equation}
After the space-time integration, we have:
\begin{eqnarray}
  \label{eq:su2-master}
  I_{R}(q,\omega)
&=&
\frac{\alpha^{2\gamma_\rho}}
     {\Gamma(\gamma_\rho+1/2)\Gamma(\gamma_\rho)\Gamma(1/2)}  
|\omega+ u_\rho
  q|^{\gamma_\rho} 
  \int_0^1 dv v^{\gamma_\rho-1/2} (1-v)^{-1/2} 
\nonumber \\ && {} \times
\frac{|\omega-(v u_\rho
    + (1-v) u_\sigma)q|^{\gamma_\rho-1}}{|u_\rho(1+v)+u_\sigma
    (1-v)|^{2\gamma_\rho}} \Theta(\omega + u_\rho q)
  \Theta (\omega
  -[v u_\rho + (1-v) u_\sigma] q),  
\end{eqnarray}
\end{widetext}
where $\Theta(x)$ is the Heaviside step function, and $\Gamma(z)$ is the
Euler Gamma function. We will consider the case of $q>0$. From
Eq.~(\ref{eq:su2-master}), we find 
 $I_{R}(q,\omega)=0$  for $\omega < u_\sigma q$
and $I_{R}(-q,-\omega)=0$  for $\omega> -u_\rho q$. Therefore,
$A_{R,s}(q,\omega)=0$ when $-u_\rho q < \omega < u_\sigma q$, while
$A_{R,s}(q,\omega)=I_{R}(q,\omega)$ for $\omega>u_\sigma q$ and
$A_{R,s}(a,\omega)=I_{R}(-q,-\omega)$ for $\omega<-u_\rho q$. For
the calculation of $I_{R}(q,\omega)$ when 
$\omega>u_\sigma q$ we have to separate the two cases, 
$u_\sigma q < \omega < u_\rho q$ and $\omega > u_\rho q$. 

\subsubsection{$\omega > u_\rho q$}

For $\omega > u_\rho q$, the two Theta functions in
Eq.~(\ref{eq:su2-master}) can be replaced by one. With the change of
variable
\begin{eqnarray}
  w=\frac{2 u_\rho  v}{u_\rho + u_\sigma + (u_\rho - u_\sigma) v}, 
\end{eqnarray}
the integral Eq.~(\ref{eq:su2-master}) becomes:
\begin{eqnarray}
  \label{eq:SU2-large-omega}
&&  A_{R,s}(q,\omega)
=\frac{  \alpha^{2\gamma_\rho}}
{
\Gamma(\gamma_\rho) \Gamma(\gamma_\rho+1)
}
\nonumber \\ && {} \quad \times
\frac{ 
(\omega+ u_\rho q)^{\gamma_\rho} (\omega - u_\sigma q)^{\gamma_\rho-1} }
{ (2u_\rho)^{\gamma_\rho+1/2}(u_\rho+u_\sigma)^{\gamma_\rho-1/2}} 
\nonumber \\ && {} \quad \times
    {}_2F_1\left(1-\gamma_\rho,\gamma_\rho+\frac{1}{2};\gamma_\rho+1;\frac{u_\rho-u_\sigma}{2u_\rho}
    \frac{\omega+u_\rho q}{\omega -u_\sigma q}\right).  
\nonumber \\
\end{eqnarray}
For $\omega \to \infty$, $I(q,\omega) \sim \omega^{2\gamma_\rho
  -1}$. {The expression (\ref{eq:SU2-large-omega}) agrees with
  Eq. (19.27) of Ref.\ \onlinecite{gogolin_1dbook} with the notation
  $\theta=2\gamma_\rho$.}  
When $\omega \to u_\rho q+0$, the argument of the hypergeometric
function in Eq.~(\ref{eq:SU2-large-omega}) becomes equal to one and
for $\gamma_\rho<1/2$, the hypergeometric function is divergent. Using
Eq.\ (\ref{eq:appF}), we
rewrite~(\ref{eq:SU2-large-omega}) as: 
\begin{eqnarray}
  A_{R,s}(q,\omega)&=&
 \frac{
    (\alpha/2u_\rho)^{2\gamma_\rho} 
    }{
\Gamma(\gamma_\rho) \Gamma(\gamma_\rho+1)
 }
\nonumber \\ &\times&
\frac{ (\omega+u_\rho q)^{\gamma_\rho} 
      (\omega - u_\rho q)^{\gamma_{\rho} -1/2} }
{(\omega  - u_\sigma q)^{1/2} }
\nonumber \\ &\times&
  {}_2F_1\left(2\gamma_\rho,\frac 1 2;\gamma_\rho+1; \frac{u_\rho
      -u_\sigma}{2 u_\rho} \frac{\omega + u_\rho q}{\omega - u_\sigma
      q}\right),   
\nonumber \\ 
\label{eq:SU2-large-omega-final} 
\end{eqnarray}
in which the hypergeometric function remains finite as $\omega \to
u_\rho q+0$. We then find that $A_{R,s}(q,\omega) \sim
\frac{\alpha^{2\gamma_\rho} q^{\gamma_\rho-1/2}
  \Gamma(1/2-\gamma_\rho) \sin(\pi \gamma_\rho)} {(2u_\rho)^{\gamma_\rho}
  (u_\rho-u_\sigma)^{1/2} \Gamma(1/2+\gamma_\rho) \pi }(\omega - u_\rho
q)^{\gamma_\rho -1/2}$ as $\omega \sim u_\rho q$. The power-law divergence was
previously obtained by Voit\cite{voit_spectral} and by Meden and
Schoenhammer\cite{meden_spectral} by analyzing the divergence of the
integral~(\ref{eq:su2-spectral}) in the vicinity of $\omega \sim
u_\rho q$.
{ The expression in terms of hypergeometric functions also
provides the prefactors.}   

\subsubsection{ $u_\sigma q < \omega< u_\rho q$}

When $u_\sigma q < \omega < u_\rho q$, the integration over $v$ in
(\ref{eq:su2-master}) is
limited to the range $0<v<(\omega - u_\sigma q)/(u_\rho q - u_\sigma
q)$. With the change of variable:
\begin{eqnarray}
  w=\frac{\omega+ u_\rho q}{\omega -u_\sigma q}
  \frac{(u_\rho-u_\sigma) v}{(u_\rho+u_\sigma) + (u_\rho-u_\sigma) v},  
\end{eqnarray}
the integral (\ref{eq:su2-master}) reduces to:
\begin{eqnarray}
&&  A_{R,s}(q,\omega)
=
\frac{\alpha^{2\gamma_\rho}}
     {\Gamma(1/2) \Gamma(2\gamma_\rho+1/2)}
\nonumber \\ && {}\quad \times
  \frac{(\omega + u_\rho q)^{-1/2}
  (\omega - u_\sigma q)^{2\gamma_\rho
    -1/2}}{(u_\rho+u_\sigma)^{\gamma_\rho-1/2} (u_\rho -
  u_\sigma)^{\gamma_\rho+1/2}} 
\nonumber \\ && {}\quad \times
{}_2 F_1 \left(\frac 1 2,
  \gamma_\rho+\frac 1 2;2\gamma_\rho + \frac 1 2; \frac{2
    u_\rho}{u_\rho - u_\sigma} \frac{\omega - u_\sigma q}{\omega +
    u_\rho q}\right).  
\nonumber \\
\label{eq:SU2-interm-omega} 
\end{eqnarray}
For $\omega \to u_\sigma q+0$, the expression
(\ref{eq:SU2-interm-omega}) has a power law divergence, $\sim (\omega
-u_\sigma q)^{2\gamma_\rho-1/2}$ for $\gamma_\rho <1/4$, in agreement
with Refs.\onlinecite{voit_spectral,meden_spectral}. When $\omega
\to u_\rho q -0$, the argument of the hypergeometric function becomes
equal to one, leading to a power law divergence. Using again Eq.\ 
(\ref{eq:appF}),  we can rewrite the
Eq.~(\ref{eq:SU2-interm-omega}) as: 
\begin{eqnarray}
&&
  A_{R,s}(q,\omega)
=
\frac{\alpha^{2\gamma_\rho}}{\Gamma(1/2)
    \Gamma(2\gamma_\rho+1/2)} 
\nonumber \\ && \quad \times
\frac{ (\omega
    - u_\sigma q)^{2\gamma_\rho-1/2} (u_\rho q
    -\omega)^{\gamma_\rho-1/2}}{  ( \omega + u_\rho q)^{\gamma_\rho}
    (u_\rho-u_\sigma)^{2\gamma_\rho}}
\nonumber \\ && \quad \times
  {}_2F_1\left(2\gamma_\rho,\gamma_\rho; 2\gamma_\rho+ 
\frac{1}{2}
; \frac{2
    u_\rho}{u_\rho - u_\sigma} \frac{\omega - u_\sigma q}{\omega +
    u_\rho q}\right)  ,
\nonumber \\
\label{eq:SU2-interm-omega-final}
\end{eqnarray}
and recover the divergence\cite{voit_spectral,meden_spectral}
$A_{R,s}\sim \frac{\alpha^{2\gamma_\rho} q^{\gamma_\rho-1/2}
  \Gamma(1/2-\gamma_\rho)} {(2u_\rho)^{\gamma_\rho}
  (u_\rho-u_\sigma)^{1/2} \Gamma(1/2+\gamma_\rho) \pi }  (u_\rho q -\omega)^{\gamma_\rho-1/2}$
for $\omega \to u_\rho q -0$. Although the exponent is the same on
both sides of $\omega = u_\rho q$, the peak is asymmetric, the ratio
of amplitudes being $\sin (\pi \gamma_\rho)$.

\subsubsection{$\omega < -u_\rho q$}

In the case $\omega < -u_\rho q$, we need
$I_{R,\sigma}(-q,-\omega)$. In the integral (\ref{eq:su2-master}),
both Theta functions are then equal to one, and with the change of
variable:
\begin{eqnarray}
  w=\frac{2 u_\rho v}{(u_\rho + u_\sigma) + (u_\rho - u_\sigma) v},
\end{eqnarray}
we find:
\begin{eqnarray}
&&
  A_{R,\sigma}(q,\omega)
=
\frac{
    \alpha^{2\gamma_\rho}
}{
\Gamma(\gamma_\rho)\Gamma(\gamma_\rho+1)
}
\nonumber \\ && {}\qquad \times
\frac{
|\omega - u_\sigma q|^{\gamma_\rho-1} 
|\omega + u_\rho q|^{\gamma_\rho}
}{
(u_\rho+u_\sigma)^{\gamma_\rho-1/2} (2 u_\rho)^{\gamma_\rho+1/2} }
\nonumber \\ && {}\qquad \times
  {}_2F_1\left(1-\gamma_\rho,\gamma_\rho+\frac{1}{2};\gamma_\rho+1;\frac{u_\rho-u_\sigma}{2 u_\rho} 
\frac{\omega+ u_\rho q}{\omega - u_\sigma q}
\right). 
\nonumber \\
\label{eq:SU2-neg-omega} 
\end{eqnarray}
For $\omega \to -u_\rho q-0$, the expression (\ref{eq:SU2-neg-omega})
vanishes with
a cusp singularity $A_{R,\sigma}(-q,-\omega) \sim
\frac{\alpha^{2\gamma_\rho} q^{\gamma_\rho-1}}{(2
  u_\rho)^{\gamma_\rho} \Gamma(\gamma_\rho) \Gamma(\gamma_\rho+1) [2
  u_\rho(u_\rho+u_\sigma)]^{1/2}} |\omega+ u_\rho
q|^{\gamma_\rho}$, in agreement with
Refs.~\onlinecite{voit_spectral,meden_spectral}.

\begin{figure}[t]
\includegraphics[width=8cm]{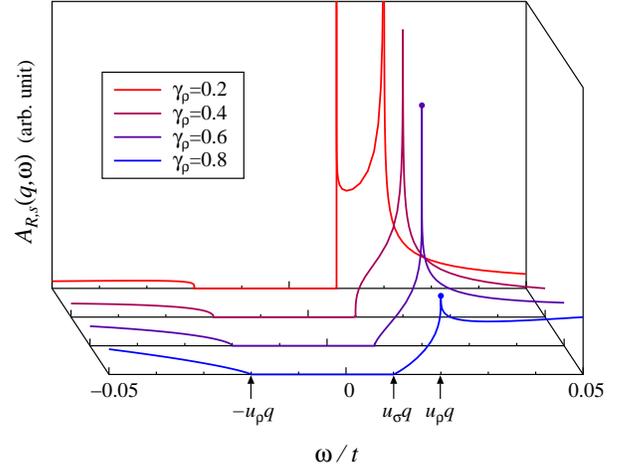}
\caption{
(Color online)
The spectral functions $A_{R,s}(q=0.01,\omega)$ with SU(2) symmetry
   for several choices of $\gamma_\rho$ 
{
($=0.2$, $0.4$, $0.6$, $0.8$ from top to bottom) 
}
  with fixed $u_\rho=2$, $u_\sigma=1$, $q=0.01$.
The dots represent the value at $\omega=u_\rho q$ given by
  Eq.\ (\ref{eq:Amax}).
}
\label{fig:spectral-SU2}
\end{figure}

Typical behavior of the spectral functions $A_{R,s}(q,\omega)$  
is  shown in Fig.\ \ref{fig:spectral-SU2}.
There is no weight at $-u_\rho q < \omega < u_\sigma q$.
The overall profile reproduces the previous results given in Refs.\
\onlinecite{meden_spectral,voit_spectral}.
We note that for $\gamma_\rho>1/2$, the divergence 
 at $\omega=u_\rho q$ disappears and is replaced by the cusp structure. 
From Eqs.\ (\ref{eq:SU2-large-omega}) and (\ref{eq:SU2-interm-omega}),
its  peak value   is given by 
\begin{equation}
A_{R,s}(q,u_\rho q) = 
\frac{\alpha^{2\gamma_\rho}
 \Gamma(\gamma_\rho-\frac{1}{2}) }
{\Gamma(\gamma_\rho)\Gamma(2\gamma_\rho)\Gamma(\frac{1}{2}) }
\frac{
(\Delta u)^{\gamma_\rho-1} q^{2\gamma_\rho-1}
}{
(2u_\rho)^{1/2}(2\bar u)^{\gamma_\rho-1/2}
}
\label{eq:Amax}
\end{equation}
where $\bar u=(u_\rho+u_\sigma)/2$ and $\Delta u=(u_\rho-u_\sigma)$.
In Fig.\ \ref{fig:spectral-SU2}, the peak positions  are
represented by the dots.
Thus the asymptotic behavior at $\omega\approx u_\rho q$ for 
$\gamma_\rho <\frac{1}{2}$ is given by
\begin{equation}
A_{R,s}
{(q,\omega)}
 = 
\left\{
\begin{array}{lll}
C (\omega-u_\rho q)^{\gamma_\rho-\frac{1}{2}} \sin \pi \gamma_\rho
&&  (\omega > u_\rho q)
\\ \\
C (u_\rho q -\omega)^{\gamma_\rho-\frac{1}{2}} & & (\omega < u_\rho q)
\end{array}
\right. ,
\end{equation}
and the asymptotic behavior for 
$\frac{1}{2}<\gamma_\rho <1$ is given by
\begin{widetext}
\begin{equation}
A_{R,s}
{(q,\omega)}
=
\left\{
\begin{array}{lll}
A_{R,s}(q,u_\rho q) 
-C'  (\omega-u_\rho q)^{\gamma_\rho-\frac{1}{2}} \sin \pi \gamma_\rho
&&  (\omega\ge u_\rho q)
\\ \\
A_{R,s}(q,u_\rho q) 
-C' (u_\rho q -\omega)^{\gamma_\rho-\frac{1}{2}} & & (\omega\le u_\rho q)
\end{array}
\right. ,
\end{equation}
\end{widetext}
where $C$ and $C'$ are positive numerical constants depending on $q$.
Because of our simplified treatment of the cutoff, 
the sum rule $\int d\omega A(q,\omega)=1$  cannot be satisfied. 
In a more rigorous treatment, the 
 short-range cutoff $\alpha$ used in the construction of the creation
 and annihilation operators\cite{haldane_bosonisation} 
and the momentum cutoff for the
 interactions must be treated
  independently. \cite{suzumura1980}  
However, in our paper, we are only concerned with the asymptotic
behavior of the spectral functions for momenta that deviate from the
Fermi momenta by an amount which is much less that the momentum
cutoff. In such a case, the corrections resulting from having
two-distinct cutoffs can be safely ignored.

\subsection{General two-band model} 
\label{sec:general-two-band}

Next we consider the case of general two-band model, and 
evaluate the integral (\ref{eq:integral}).
By using twice the Feynman identity (\ref{eq:feynman}) and
Eq.~(\ref{eq:transformed}),  
we can reexpress Eq.\ (\ref{eq:integral}) as
\begin{widetext}
  \begin{eqnarray} \label{eq:master}
    I_{R,a}(q,\omega)&=&\frac{
     \alpha^{\bar \nu_{a}-1}}
  {\Gamma(\nu_{a,+})\Gamma(\nu_{a,-})\Gamma(\nu'_{a,+})\Gamma(\nu'_{a,-})}
     \int_0^1 dw_1 \int_0^1 dw_2  \, 
w_1^{\nu_{a,+}-1} (1-w_1)^{\nu_{a,-}-1} 
w_2^{\nu'_{a,+}-1}   (1-w_2)^{\nu'_{a,-} -1} 
\nonumber \\ && \times
[\omega - u(w_1) q]^{\nu'_{a,+}+\nu'_{a,-}-1} 
[\omega + u(w_2) q ]^{\nu_{a,+}+\nu_{a,-}-1}
 \frac{\Theta[\omega -u(w_1) q] \Theta[\omega + u(w_2) q]}
      {[u(w_1)+u(w_2)]^{\bar \nu_{a}-1}} ,
  \end{eqnarray}
\end{widetext}
where $u(w)\equiv w u_+  + (1-w) u_-$.
From  Eq.\ (\ref{eq:master}), we obtain  
$A_{R,a}(q>0,|\omega|<u_- q) =0$. Contrarily to the SU(2) symmetric
case of Sec.~\ref{sec:su2}, the spectral function does not vanish
anymore when $-u_+ q < \omega < -u_- q$.

\subsubsection{Power-law singularities} 
The power-law
singularities\cite{meden_spectral,voit_spectral} can be recovered from
Eq.\ (\ref{eq:master}) in a simple manner. 
In the analysis of the power-law singularities, 
 we restrict ourselves to the case $q>0$.
Indeed, if we consider the case
of $\omega \to u_+ q+0$, the dominant contribution to the integral
comes from the integration over $w_1$ in the vicinity of $w_1=1$. 
The integral to consider is then:
\begin{eqnarray}
    I_{R,a}(q,\omega)&\propto&
  \int_0^1 dw_1 (1-w_1)^{\nu_{a,-} -1} 
\nonumber \\ && {} \times
[(\omega - u_+ q) + 
\Delta u
q
  (1-w_1)]^{\nu'_{a,-}+\nu'_{a,+}-1} 
\nonumber \\ 
&\approx& |\omega - u_+ q
  |^{\nu_{a,-}+\nu'_{a,-}+\nu'_{a,+}-1} |\Delta u q|^{-\nu_{a,-}}, 
\nonumber \\ &&    
  \label{eq:approx-w-uplusq}
\end{eqnarray}
provided that $\nu_{a,-}+\nu'_{a,-}+\nu'_{a,+}<1$, 
 where $\Delta u\equiv (u_+-u_-)$.
Similarly, when $u_-q < \omega < u_+q$, the $w_1$ integration also
determines the power law singularities. This time, the integration
over $w_1$ is restricted to 
$0<w_1<(\omega-u_-q)/(\Delta u q)$. 
Changing variables to 
$\bar{w}_1= w_1 (\omega-u_-q)/(\Delta u q)$, 
we have to consider the integral: 
\begin{eqnarray}
&&
  \frac{(\omega-u_-q)^{\nu_{a,+}+\nu'_{a,-}+\nu'_{a,+}-1}}
 {(\Delta uq)^{\nu_{a,+}-1}}
   \int_0^1 d\bar{w}_1 \bar{w}_1^{\nu_{a,+}-1}
\nonumber \\ && {} \quad \times
 (1-\bar{w}_1
)^{\nu'_{a,-}+\nu'_{a,+}-1} \left(1-\frac{\omega - u_- q}{\Delta u q} \bar{w}_1\right)^{\nu_{a,-}-1}.
\nonumber \\ 
\end{eqnarray}
When $\omega \to u_- q+0$, the integral goes to a constant, and we
have the power 
law: $I_{R,a}(q,\omega) \sim
(\omega-u_-q)^{\nu_{a,+}+\nu'_{a,+}+\nu'_{a,-}-1}$. When $\omega \to
u_+ q -0$, the integral has a power law singularity, and the behavior
$I_{R,a}(q,\omega) \sim
  (-\omega+u_+q)^{\nu_{a,-}+\nu'_{a,+}+\nu'_{a,-}-1} $
is obtained. Such
behavior was previously obtained in
Refs.\ \onlinecite{meden_spectral,voit_spectral}. 
By a similar method, one can also obtain the power law singularities
at $\omega=-u_\pm q$ of $I_{R,a}(-q,-\omega)$. This time, the origin
of the singularities is the integration over $w_2$. 
By summarizing, the asymptotic behavior of
the spectral function are given by
\begin{equation}
A_{R,a}(q,\omega)
\propto
\left\{
\begin{array}{ll}
|\omega- u_+ q|^{\beta_{a,+}}  & \mbox{(for $\omega \to + u_+ q \pm 0$)}
\\
(\omega- u_- q)^{\beta_{a,-}}  & \mbox{(for $\omega \to + {u_-} q +0$)}
\\
(\omega + u_- q)^{\beta'_{a,-}}  & \mbox{(for $\omega \to -u_- q -0$)}
\\ C+
|\omega + u_+ q|^{\beta'_{a,+}}  & \mbox{(for $\omega \to -u_+ q \pm 0$)}
\end{array}
\right.,
\end{equation}
where
\begin{subequations}
\label{eq:exponent_beta}
\begin{eqnarray}
\beta_{a,+} &\equiv& \nu_{a,-} + \nu'_{a,+} + \nu'_{a,-} - 1, \\
\beta_{a,-} &\equiv& \nu_{a,+} + \nu'_{a,+} + \nu'_{a,-} - 1, \\
\beta'_{a,-} &\equiv& \nu_{a,+} + \nu_{a,-} + \nu'_{a,+} - 1, \\
\beta'_{a,+} &\equiv& \nu_{a,+} + \nu_{a,-} + \nu'_{a,-} - 1, 
\end{eqnarray}
\end{subequations}
As
$\nu_{a,+}+\nu_{a,-}=1+\nu'_{a,+}+\nu'_{a,-}$, there is no divergence
but only a cusp in the vicinity of  $\omega = - u_{\pm} q$.

From Eq.\ (\ref{eq:exponent}), 
the exponents which determine the Green's function for the right moving
particle in Eqs.~(\ref{eq:integral}) are given by 
\begin{eqnarray}
\nu_{1,\pm} &=& 
\frac{1}{8}
\left(
\frac{K_1u_1}{u_\pm} + \frac{u_\pm}{K_1u_1} +  2 
\right)
\left(
1\pm\frac{u_1^2-u_2^2}{u_+^2-u_-^2}
\right),
\nonumber \\
\nu'_{1,\pm} &=&
\frac{1}{8}
\left(
\frac{K_1u_1}{u_\pm} + \frac{u_\pm}{K_1u_1} -  2 
\right)
\left(
1\pm\frac{u_1^2-u_2^2}{u_+^2-u_-^2}
\right),
\nonumber \\
\nu_{2,\pm} &=& 
\frac{1}{8}
\left(
\frac{K_2u_2}{u_\pm} + \frac{u_\pm}{K_2u_2} +  2 
\right)
\left(
1\mp \frac{u_1^2-u_2^2}{u_+^2-u_-^2}
\right),
\nonumber \\
\nu'_{2,\pm} &=&
\frac{1}{8}
\left(
\frac{K_2u_2}{u_\pm} + \frac{u_\pm}{K_2u_2} -  2 
\right)
\left(
1\mp \frac{u_1^2-u_2^2}{u_+^2-u_-^2}
\right).
\nonumber \\
\label{eq:mu}
\end{eqnarray}  

\begin{figure}[t]
\includegraphics[width=6cm]{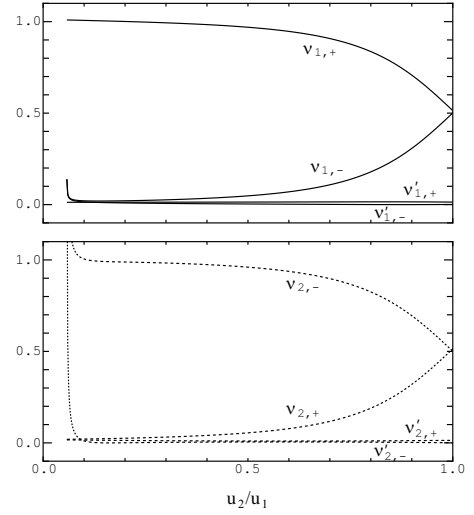}
\caption{
The $u_2/u_1$ dependence of the exponents 
$\nu_{1,\pm}$, $\nu'_{1,\pm}$ (top) and 
$\nu_{2,\pm}$, $\nu'_{2,\pm}$ (bottom)
for  $K_1=K_2=0.8$, and  $g/(\pi u_1)=0.3$.
The two-component TL liquid is unstable 
  for $u_2 < u_{2c} \approx 0.0576 u_1$.
}
\label{fig:Fig4N1}
\end{figure}
\begin{figure}[t]
\includegraphics[width=6cm]{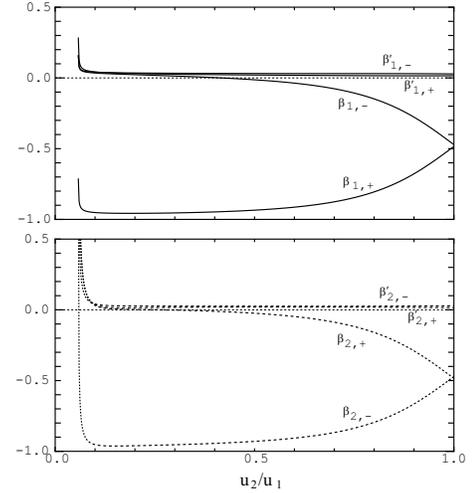}
\caption{
The $u_2/u_1$ dependence of the exponents 
$\beta_{1,\pm}$, $\beta'_{1,\pm}$ (top) and 
$\beta_{2,\pm}$, $\beta'_{2,\pm}$ (bottom)
for  $K_1=K_2=0.8$, and  $g/(\pi u_1)=0.3$.
}
\label{fig:Fig4N2}
\end{figure}

Figure~\ref{fig:Fig4N1} shows 
the exponents $\nu_{a,\beta}$ [Eq.~(\ref{eq:mu})]
  as a function of difference in velocities.
In order to distinguish the effects of the 
intraband coupling $g_a$ and the interband coupling $g$,
 we plot the $u_2/u_1$ dependence of the exponents with fixed 
 $K_a$ (i.e., fixed $g_a/v_a$).
In this case, the  two-component TL liquid 
  is always unstable for small $u_2$,
  in contrast to the situation in Fig.\ \ref{fig:phasediagram-v1v2}.
 For $u_1=u_2$, the behavior is similar to 
that of the TL liquid  with only the intraband forward scattering 
where
 $(\nu_{a,+},\nu_{a,-},\nu'_{a,+},\nu'_{a,-})
  \approx (0.514,0.501,0.014,0.001)$,
suggesting the weak effect of the intraband interactions.
With decreasing $u_2$,
the exponent  $\nu_{1,-}$ ($\nu_{2,+}$) decreases 
suggesting that the integral at $x\approx u_- t$ ($x\approx u_+ t$)
for $I_{R,1}$ ($I_{R,2}$)  in Eq.\ (\ref{eq:integral})
  becomes less singular.
Although the effect of intraband interaction on  
$\nu'_{1,\pm}$ and $\nu'_{2,\pm}$ 
is small for $u_2/u_1 \simeq 1$, 
the exponents
  $\nu'_{1,-}$ and  $\nu'_{2,-}$ 
 are enhanced  for small $u_2$ just above  $u_{2c}$.
 In Fig.~\ref{fig:Fig4N2},  the corresponding 
  $\beta_{1,\pm}$, $\beta_{2,\pm}$,  
   $\beta'_{1,\pm}$, and $\beta'_{2,\pm}$ 
are shown. 
The case of $K=1$ (not shown) is similar to Fig.~\ref{fig:Fig4N2}.
Note that
the exponent $\nu'_{1,+}$ becomes extremely small but nonzero
for $K_1=K_2=1$ and small $g/\pi$. 
Its asymptotic behavior is given by
$\nu'_{1,+} \simeq [g/(\pi u_1)]^4(u_2/u_1)^2/16$
for $u_2 \ll u_1$.

\begin{figure}[t]
\includegraphics[width=6cm]{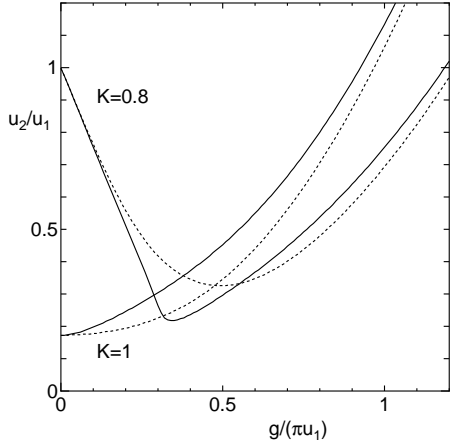}
\caption{
The values
for $u_2/u_1$ satisfying
$\beta_{2,+}=0$ (solid line) and $\beta_{1,-}=0$ (dotted
line), with fixed $K(=K_1=K_2)=1$ and $0.8$.
}
\label{fig:critical_value}
\end{figure}

As will be shown in the next section, 
the spectral function  has two divergences
 at $\omega = u_+q$ and $\omega = u_-q$ 
 when  $\beta_{2,+}<0$ and $\beta_{1,-}<0$, respectively. 
However each divergence is replaced by a cusp when 
 $\beta_{2,+}>0$ and $\beta_{1,-}>0$, respectively. 
Figure~\ref{fig:critical_value} shows such a critical value as a function 
 of  $g/(\pi u_1)$ with fixed $\beta_{2,+}=0$ (solid lines)
 and $\beta_{1,-}=0$ (dotted lines)
 for two cases where 
  the intraband interaction is absent ($K_1=K_2=1$) 
   and is present  ($K_1=K_2=0.8$).
For $K_a=1$, the critical value of $u_2$ increases as a function of $g$ 
 while  that takes a minimum for $K=0.8$. 
The minimum for $K \not= 1$  suggests a competition 
  between the intraband interaction and the interband interaction. 
With increasing   $K$ to 1, the value of  $g/(\pi  u_1)$ at which $u_2$ takes a
 minimum decreases, 
and the line in the limit of $K \rightarrow 1$ 
  coincides with that of $K=1$ except 
 for $g/\pi=0$, where  $u_2/u_1=1$ at $g/\pi=0$ 
 is always obtained for $K \not=0$.
 Thus it is found that double peaks of either  1 or 2 particle 
 is replaced by  a single peak 
   in the region enclosed by the solid line and the dotted line
  in Fig.~\ref{fig:critical_value}.

\subsubsection{Representation of spectral function as integrals of hypergeometric functions}
\label{sec:hypergeo}

It is possible to rewrite the double integral~(\ref{eq:master}) as a
single integral containing the
 Gauss hypergeometric function\cite{abramowitz_math_functions}
 ${}_2F_1(\alpha,\beta;\gamma;z)$
(see Appendix \ref{sec-app-integral}).
We again restrict ourselves to the case of $q>0$. 
 The spectral function for $q<0$ can easily obtained by noting 
 $A_{\nu,a}(q,\omega)=A_{\nu,a}(-q,-\omega)$ [see Eq.\ (\ref{eq:AtoI})].
For $|\omega| > u_+ q$, we obtain
\begin{widetext}
\begin{subequations}
\begin{eqnarray}
   A_{R,a}(q,\omega)|_{\omega>u_+q}
&=&
\frac{ 
(q\alpha)^{\bar \nu_{a}-1}
}
  {\Gamma(\nu_{a,+}+\nu_{a,-})\Gamma(\nu'_{a,+})\Gamma(\nu'_{a,-})}
\frac{
  (\omega-u_+q)^{\nu_{a,-}+\nu_{a,+}'+\nu_{a,-}'-1} 
  (\omega + u_+q)^{\nu_{a,+}+\nu_{a,-}-1}
}
{
  (\omega - u_- q)^{\nu_{a,-}}
 (2u_+ q)^{\nu_{a,+}+\nu_{a,-}+\nu'_{a,+}-1}  (2\bar u q)^{\nu'_{a,-}}
}
\nonumber \\ &&  
\times  
\int_0^1 dt \,\,
\left( 1-t \right)^{\nu'_{a,+}-1} 
\left( t \right)^{\nu'_{a,-} -1}
\left[
1+
\frac{\Delta u (\omega - u_+q)}{2\bar u  (\omega + u_+q)}
t
\right]^{\nu_{a,+}+\nu_{a,-}-1}
\nonumber \\ && {} \times
\, {}_2 F_1 \left(\bar\nu_{a}-1, \, \nu_{a,-}; \,\nu_{a,+}+\nu_{a,-}; \,
z_1
\right),
\label{eq:A1}
\\
   A_{R,a}(q,\omega)|_{\omega<-u_+q}
&=&
\frac{ 
(q\alpha)^{\bar \nu_{a}-1}
}
  {\Gamma(\nu'_{a,+}+\nu'_{a,-})\Gamma(\nu_{a,+})\Gamma(\nu_{a,-})}
\frac{
  (|\omega|-u_+q)^{\nu_{a,+}+\nu_{a,-}+\nu'_{a,-}-1} 
  (|\omega| + u_+q)^{\nu'_{a,+}+\nu'_{a,-}-1}
}
{
  (|\omega| - u_- q)^{\nu'_{a,-}}
 (2u_+ q)^{\nu_{a,+}+\nu'_{a,+}+\nu'_{a,-}-1}  (2\bar u q)^{\nu_{a,-}}
}
\nonumber \\ &&  
\times  
\int_0^1 dt \,\,
\left( 1-t \right)^{\nu_{a,+}-1} 
\left( t \right)^{\nu_{a,-} -1}
\left[
1+
\frac{\Delta u (|\omega| - u_+q)}{2\bar u  (|\omega| + u_+q)}
t
\right]^{\nu'_{a,+}+\nu'_{a,-}-1}
\nonumber \\ && {} \times
\, {}_2 F_1 \left(\bar\nu_{a}-1, \, \nu'_{a,-}; \,\nu'_{a,+}+\nu'_{a,-}; \,
z_1
\right).
\label{eq:A4}
\end{eqnarray}%
\label{eq:A1A4}%
\end{subequations}
Here $\bar u \equiv (u_+ + u_-)/2 $ 
{and} $\Delta u \equiv  (u_+ - u_-)$.
The parameter $z_1$ is given by
\begin{eqnarray}
z_1=
\frac{\Delta u (|\omega| + u_+q)}{ 2u_+ (|\omega| - u_- q)}
\left[
1+
\frac{\Delta u (|\omega| - u_+q)}{2\bar u (|\omega| + u_+q)}
t
\right].
\end{eqnarray}
For $u_- q < |\omega| < u_+ q$, we have:
\begin{subequations}
\begin{eqnarray}
A_{R,a}(q,\omega)|_{u_- q < \omega < u_+ q}
&=&
\frac{ 
  (q\alpha)^{\bar \nu_{a}-1}
}
{\Gamma(\nu_{a,+}+\nu'_{a,+}+\nu'_{a,-})\Gamma(\nu_{a,-})
 B(\nu'_{a,+},\nu'_{a,-})
}
\nonumber \\ && {}\times
\frac{(\omega - u_- q) ^{\nu_{a,+}+\nu'_{a,+}+\nu'_{a,-}-1} 
 (- \omega + u_+ q)^{\nu_{a,-}+\nu'_{a,-}+\nu'_{a,+}-1} 
}
{(\omega + u_+ q)^{\nu'_{a,+}} (\omega+u_-q)^{\nu'_{a,-}} 
 (\Delta uq)^{\bar \nu_{a}-1} 
}
\nonumber \\ && {}\times
 \int_0^1 dt \,\,
 (1-t)^{\nu'_{a,+}-1} (t)^{\nu'_{a,-} -1}
\, {}_2 F_1 \left(\bar \nu_{a}-1, \,
	     \nu'_{a,+}+\nu'_{a,-}; \, \nu_{a,+}+\nu'_{a,+}+\nu'_{a,-}; \,
z_2
\right),
\label{eq:A2}
\\
A_{R,a}(q,\omega)|_{-u_+ q < \omega < -u_- q}
&=&
\frac{ 
  (q\alpha)^{\bar \nu_{a}-1}
}
{\Gamma(\nu_{a,+}+\nu_{a,-}+\nu'_{a,+})
\Gamma(\nu'_{a,-})
 B(\nu_{a,+},\nu_{a,-})
}
\nonumber \\ && {}\times
\frac
{(|\omega| - u_- q) ^{\nu_{a,+}+\nu_{a,-}+\nu'_{a,+}-1} 
 (-|\omega| + u_+ q)^{\nu_{a,+}+\nu_{a,-}+\nu'_{a,-}-1} 
}
{(|\omega| + u_+ q)^{\nu_{a,+}} (|\omega|+u_-q)^{\nu_{a,-}} 
 (\Delta uq)^{\bar \nu_{a}-1} 
}
\nonumber \\ && {}\times
 \int_0^1 dt \,\,
 (1-t)^{\nu_{a,+}-1} (t)^{\nu_{a,-} -1}
\, {}_2 F_1 \left(\bar \nu_{a}-1, \,\nu_{a,+}+\nu_{a,-}; \, 
                 \nu_{a,+}+\nu_{a,-}+\nu'_{a,+}; \, z_2
\right), \quad\qquad
\label{eq:A3}
\end{eqnarray}%
\label{eq:A2A3}%
\end{subequations}
\end{widetext}
where $B(p,q)$ is the beta function
 $B(p,q)=\Gamma(p)\Gamma(q)/\Gamma(p+q)$.
The parameter $z_2$ is given by
\begin{eqnarray}
z_2 =
\frac{ 2  u_+ (|\omega| - u_- q)}{\Delta u(|\omega|+u_+q)} 
\left[
1
+
\frac{\Delta u (-|\omega|  +  u_+q)}{ 2  u_+ (|\omega|+u_-q)} t
\right].
\end{eqnarray}
We find 
$A_{R,a}(q,\omega)|_{-u_-q < \omega < u_- q}=0$.

We note that, if $\nu'_{a,-}=0$, we have $A_{R,a}=0$ for
  $-u_+ q < \omega < -u_- q<0$.
In Eq.\ (\ref{eq:A1A4}),
 the power-law singularities are already
factored out and 
the integrals are regular.
 This fact allows the numerical evaluation of the formulae
 (\ref{eq:A1A4}) and (\ref{eq:A2A3}).

The integrals in Eq.\ (\ref{eq:A1A4}) can also be reduced to 
  the Appell hypergeometric functions\cite{erdelyi_functions_1,dlmf_appell}. By using 
Eq.\ (\ref{eq:AppellF1}), 
the spectral functions 
for the range $u_- q < |\omega| < u_+ q$ are expressed as
\begin{widetext}
\begin{eqnarray}
A_{R,a}(q,\omega)|_{u_- q < \omega < u_+ q}
&=&
\frac{ (\alpha/\Delta u)^{\bar \nu_{a}-1}}
{\Gamma(\nu_{a,+}+\nu'_{a,+}+\nu'_{a,-})\Gamma(\nu_{a,-})
}
\frac{(|\omega| - u_- q) ^{\nu_{a,+}+\nu'_{a,-}+\nu'_{a,+}-1} 
 (- |\omega| + u_+ q)^{\nu_{a,-}+\nu'_{a,-}+\nu'_{a,+}-1} 
}
{(|\omega| + u_+ q)^{\nu'_{a,+}} (|\omega|+u_-q)^{\nu'_{a,-}} 
}
\nonumber \\ && {}\times
F_1 \left(\bar \nu_{a}-1 ; \nu'_{a,+},\nu'_{a,-}
     ;\nu_{a,+}+\nu'_{a,+}+\nu'_{a,-}; 
\frac{ 2  u_+ (|\omega| - u_- q)}{\Delta u(|\omega|+u_+q)} 
 , \frac{2\bar{u}(|\omega| - u_- q)}{\Delta u(|\omega|+u_-q)} \right),
\label{eq:A2-final} 
\\ 
A_{R,a}(q,\omega)|_{-u_+ q < \omega < -u_- q}
&=&
\frac{ 
  (\alpha/\Delta u)^{\bar \nu_{a}-1}
}
{\Gamma(\nu_{a,+}+\nu_{a,-}+\nu'_{a,+})
\Gamma(\nu'_{a,-})
}
\frac
{(|\omega| - u_- q) ^{\nu_{a,+}+\nu_{a,-}+\nu'_{a,+}-1} 
 (-|\omega| + u_+ q)^{\nu_{a,+}+\nu_{a,-}+\nu'_{a,-}-1} 
}
{(|\omega| + u_+ q)^{\nu_{a,+}} (|\omega|+u_-q)^{\nu_{a,-}} 
}
\nonumber \\ && {}\times
F_1 \left(\bar \nu_{a}-1; \nu_{a,+},\nu_{a,-};
  \nu_{a,+}+\nu_{a,-}+\nu'_{a,+}; 
\frac{ 2  u_+ (|\omega| - u_- q)}{\Delta u(|\omega|+u_+q)} 
 , \frac{2\bar{u}(|\omega| - u_- q)}{\Delta u(|\omega|+u_-q)} 
\right),
\end{eqnarray}
where 
$F_1(\alpha;\beta,\beta' ; \gamma;  x,  y)$ 
is the first Appell hypergeometric function of two variables.
Similarly, 
by using Eq.\ (\ref{eq:AppellF2}), 
 equation (\ref{eq:A2A3}) can also be reduced to 
\begin{eqnarray}
   A_{R,a}(q,\omega)|_{\omega > u_+ q}
&=&
\frac{ 
(\alpha/2u_+)^{\bar \nu_{a}-1}
}
  {\Gamma(\nu_{a,+}+\nu_{a,-})\Gamma(\nu'_{a,+}+\nu'_{a,-})}
\frac{
  (|\omega|-u_+q)^{\nu_{a,-}+\nu_{a,+}'+\nu_{a,-}'-1} 
  (|\omega| + u_+q)^{\nu_{a,+}+\nu_{a,-}+\nu'_{a,-}-1}
}
{
  (|\omega| - u_- q)^{\nu_{a,-}} (|\omega| + u_-q)^{\nu'_{a,-}}
}
\nonumber \\ &&  
\times  
 F_2 \left(\bar\nu_{a}-1; \, \nu_{a,-}, \, \nu'_{a,-},  \,
   \nu_{a,+}+\nu_{a,-}, \,
\nu'_{a,+}+\nu'_{a,-} ; \,
\frac{\Delta u (|\omega| + u_+q)}{ 2u_+ (|\omega| - u_- q)}
, \,
\frac{\Delta u (|\omega| - u_+q)}{2u_+ (|\omega| + u_-q)}
\right),
\qquad
\label{eq:A1-final}
\\ 
   A_{R,a}(q,\omega)|_{\omega < -u_+ q}
&=&
\frac{ 
(\alpha/2u_+)^{\bar \nu_{a}-1}
}
  {\Gamma(\nu'_{a,+}+\nu'_{a,-})\Gamma(\nu_{a,+}+\nu_{a,-})}
\frac{
  (|\omega|-u_+q)^{\nu_{a,+}+\nu_{a,-}+\nu'_{a,-}-1} 
  (|\omega| + u_+q)^{\nu_{a,-}+\nu'_{a,+}+\nu'_{a,-}-1}
}
{
  (|\omega| - u_- q)^{\nu'_{a,-}} (|\omega| + u_-q)^{\nu_{a,-}}
}
\nonumber \\ &&  
\times  
 F_2 \left(\bar\nu_{a}-1; \, \nu'_{a,-} ,  \, \nu_{a,-} ;
    \,\nu'_{a,+}+\nu'_{a,-}, \,
\nu_{a,+}+\nu_{a,-} ;  \,
\frac{\Delta u (|\omega| + u_+q)}{ 2u_+ (|\omega| - u_- q)}
, \,
\frac{\Delta u (|\omega| - u_+q)}{2u_+ (|\omega| + u_-q)}
\right),
\label{eq:A4-final}
\end{eqnarray}
\end{widetext}
where 
$F_{{2}}(\alpha;\beta,\beta' ; \gamma, \gamma';  x,  y)$ is
the second Appell hypergeometric function\cite{erdelyi_functions_1,dlmf_appell}.
These results are
reminiscent of the one from Ref.\ \onlinecite{iucci2007}, where the Fourier
transform of the $2k_F$ component of the density-density correlation
function of a TL liquid with different spin and charge
velocities was shown to be expressible in terms of the  Appell
hypergeometric function of two variables. 
We note also that by setting
$u_+ = u_\rho$, $u_-= u_\sigma$, $\nu_{a,+}=(\gamma_\rho+1/2)$, 
  $\nu_{a,-}=1/2$, $\nu_{a,+}'=\gamma_\rho$, and $\nu_{a,-}'=0$
in Eqs.\ (\ref{eq:A2-final}), (\ref{eq:A1-final}), and (\ref{eq:A4-final}),
we can reproduce Eqs.\ (\ref{eq:SU2-interm-omega-final}), 
(\ref{eq:SU2-large-omega-final}), and (\ref{eq:SU2-neg-omega}), 
 respectively, showing that the Appell hypergeometric function
 representation also covers the SU(2) invariant case.

The special case of
SU(2) symmetry can also be recovered from the integral representations
(\ref{eq:A1A4}) and (\ref{eq:A2A3}). 
However, to do this, we first need to consider the limit  $\nu_2 \to 0$ in
Eq.~(\ref{eq:feynman}). 
Rewriting~(\ref{eq:feynman}) as:
\begin{eqnarray}
 \frac{1}{A_1^{\nu_1} A_2^{\nu_2}}
    &=&\frac{\Gamma(\nu_1+\nu_2)}{\Gamma(\nu_1) \Gamma(\nu_2+1)} 
\nonumber \\ &\times&
\int_0^1
    dw \frac{ \nu_2 w^{\nu_2-1} (1-w)^{\nu_1-1}}{\left[A_1 (1-w) +
        A_2w \right]^{\nu_1+\nu_2}}, 
\end{eqnarray}
by  making $w \to (1-w)$ and using $\Gamma(\nu_2+1)=\nu_2
\Gamma(\nu_2)$ and changing integration variable to {$v=w^{\nu_2}$},
we find that:
\begin{eqnarray}
  \frac{1}{A_1^{\nu_1} A_2^{\nu_2}}
    &=&\frac{\Gamma(\nu_1+\nu_2)}{\Gamma(\nu_1) \Gamma(\nu_2+1)} 
\nonumber \\ &\times&
\int_0^1
    dv \frac{ (1-v^{1/\nu_2})^{\nu_1-1}}{\left[A_1 (1-v^{1/\nu_2}) +
        A_2 v^{1/\nu_2} \right]^{\nu_1+\nu_2}}, 
\nonumber \\
\end{eqnarray}
With this form, when $\nu_2 \to 0$, $v^{1/\nu_2} \to 0
\forall v<1$, and the integral reduces to $A_1^{-\nu_1}$ so that the
Feynman identity remains applicable when $\nu_2 \to 0$.  
By applying the same transformations to Eq.~(\ref{eq:A1}) , we see 
that  in the limit
of $\nu'_{a-}\to 0$, the  SU(2)
case is recovered. The same result also applies to Eq. (\ref{eq:A2}). In
Eq. (\ref{eq:A4}), when $\nu'_{a+} \to 0$, 
the hypergeometric function in the integrand reduces to $1$. 
The integral in Eq. (\ref{eq:A4}) thus becomes a hypergeometric
function ${}_2F_1$ and the SU(2) invariant case is again recovered. 
Finally, 
in Eq. (\ref{eq:A3}), a factor $\Gamma(\nu'_{a,-})$ 
is present in the denominator of
the fraction, but the integral remains finite in the limit
$\nu'_{a,-} \to 0$. As a result, in the limit $\nu'_{a,+}\to 0$, the
contribution of Eq.~(\ref{eq:A3}) to the spectral function vanishes, 
recovering fully the SU(2) invariant result.

\subsubsection{Application to the Hubbard model} 
\label{sec:hubbard-model}

\begin{figure}[t]
\includegraphics[width=7cm]{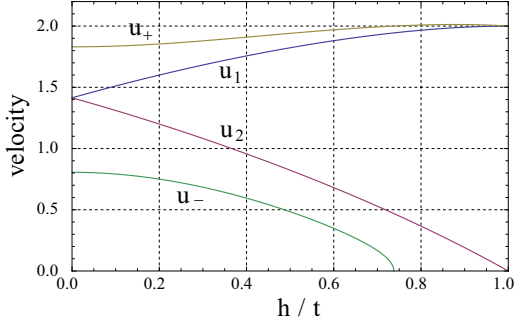}
\caption{
(Color online)
Velocities as a function of the magnetic field
for 
$U/t=3$ and $\rho=1/2$.
}
\label{fig:velocity}
\end{figure}

\begin{figure}[t]
\includegraphics[width=8cm]{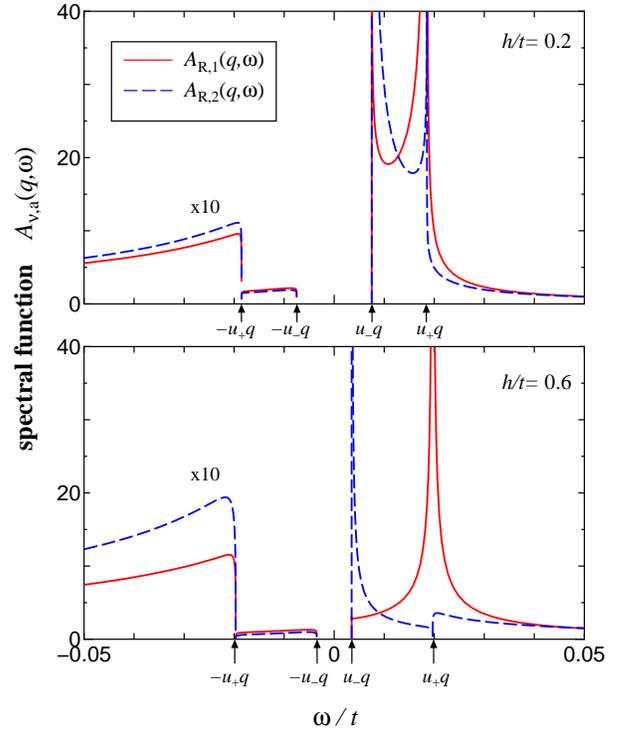}
\caption{
(Color online)
The spectral functions $A_{R,1}(q,\omega)$ 
  and $A_{R,2}(q,\omega)$ for $U/t=3$ and $q=0.01/a$, 
 with fixed $h/t=0.2$ (top) and 0.6 (bottom).
}
\label{fig:spectral}
\end{figure}

\begin{figure*}[t]
\includegraphics[width=14cm]{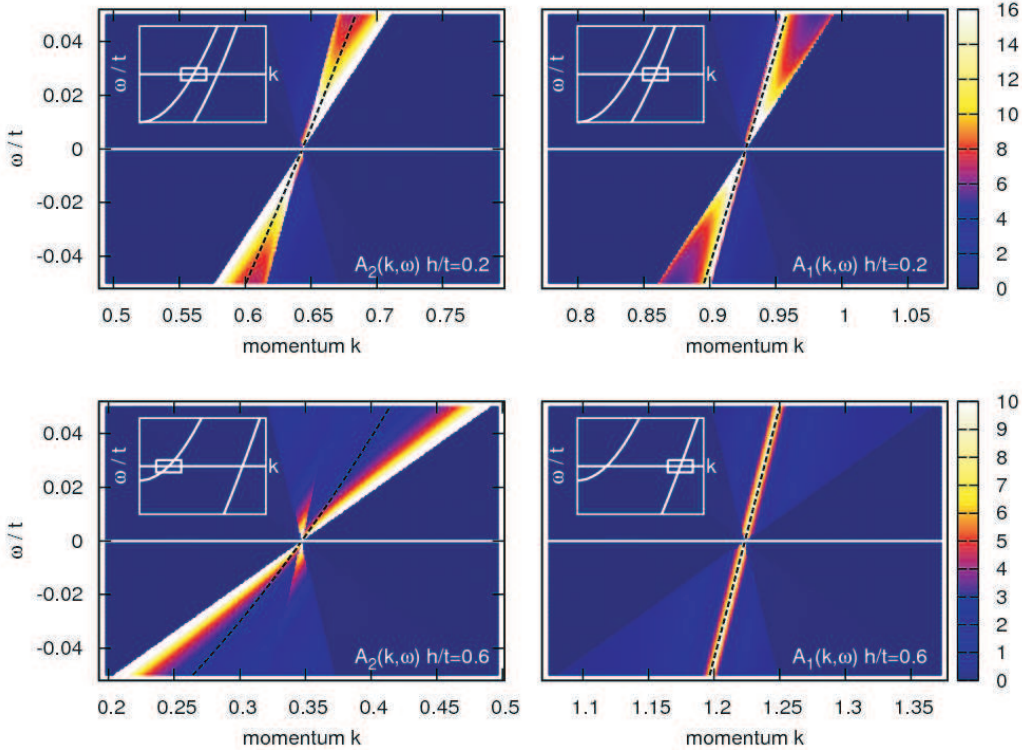}
\caption{
(Color online)
Contour plot of the spectral function $A_a(k,\omega)$ for $a=2$
 (left) and $a=1$ (right), with fixed $U/t=3$ and $h/t=0.2$ (top) 
 and $h/t=0.6$ (bottom).
The dotted lines denote the bare dispersion 
{$\varepsilon_a(k) \simeq u_a(k-k_{F,a})$}
where the Fermi momenta are given by
 $(k_{F,2},k_{F,1})\simeq (0.64,0.93)$ for $h/t=0.2$ and
 $(0.35,1.22)$ for $h/t=0.6$.
In the insets, 
the bare energy dispersions are shown and the 
 rectangular represent the region where the spectral functions 
  are plotted in the main figures. 
}
\label{fig:contour}
\end{figure*}

For the application of the present calculation to an explicit model, 
here we consider the spin-$\frac{1}{2}$
Hubbard model under the magnetic field:
\begin{eqnarray}
 H&=&-t\sum_{j,\sigma}(c_{j,\sigma}^\dagger
  c_{j+1,\sigma}+\mathrm{h.c.})
-\mu \sum_{j,\sigma} n_{j,\sigma}
\nonumber \\ && {}
{
-h \sum_{j} (n_{j,\uparrow}- n_{j,\downarrow})
}
+U\sum_j n_{j,\uparrow} n_{j,\downarrow} ,
\label{eq:Hubbard}
\end{eqnarray}
where 
{$\sigma=\uparrow,\downarrow$} refers to the spin degrees of freedom.
We restrict ourselves to the case of repulsive interaction ($U>0$).
Here we assume that 
$h>0$ and the energy dispersion is given by
 $\varepsilon_{a=1(2)}(k)=-2t\cos k -(+)h -\mu$,
where $a=1$ ($a=2$) corresponds 
 to the majority (minority) spin.

We denote the filling factor by $\rho$ ($0 \le \rho \le 2$).
For  $h<t(1-\cos \pi \rho)$,
two bands are overlapped at Fermi energy. 
The Fermi momenta and
the bare velocities are given by
\begin{eqnarray}
k_{F,1(2)}&=&
\frac{\pi\rho}{2} +(-) 
\sin^{-1}\left(\frac{h}{2t\sin \frac{\pi\rho}{2}}\right),
\\
u_{1(2)}&=&
2ta
\sin\frac{\pi\rho}{2}\sqrt{
1-\frac{h^2}{4t^2 \sin^2 \frac{\pi\rho}{2}}
}
\nonumber \\ && {}
+ (-) 2ta
\cos\frac{\pi\rho}{2}
\frac{h}{2t \sin \frac{\pi\rho}{2}}.
\end{eqnarray}  
The effective velocities $u_\pm$ are given by Eq.\ (\ref{eq:velocity}) 
 with $g=U$ and $K_1=K_2=1$.
By considering incommensurate band filling,
  the umklapp scattering is irrelevant.
For $h=0$, the Hubbard Hamiltonian (\ref{eq:Hubbard}) possesses 
 the SU(2) symmetry and the analysis given in Sec.\ \ref{sec:su2} 
  applies.
The neglect of the backward scattering can be justified
 for strong magnetic field, and 
  the effective model can be described by Eq.\ (\ref{eq:hamiltonian})
  with $g=U$.
The magnetic field dependence of the bare velocities $u_{1,2}$ and the
  renormalized velocities $u_\pm$
 are
 shown in Fig.\ \ref{fig:velocity}.
For $h=0$, the velocities $u_+$ and $u_-$ correspond
  to the conventional charge $v_\rho$ and spin $v_\sigma$ excitation
  velocities,
respectively.

The spectral functions $A_{R,a}(q,\omega)$  for 
 $h/t=0.2$ and $0.6$ are shown in Fig.\ \ref{fig:spectral}.
In Fig.\ \ref{fig:contour},
the contour plot of $A_a(k,\omega)$ is
  shown in the range $-0.05<(k-k_{F,a})<0.05$ and 
  $-0.05<\omega/t<0.05$, with fixed $h/t=0.2$ (top figures) and $0.6$
  (bottom figures).
The momentum $k$ is related to $q$ by 
$k=k_{F,a}+q$ [see Eq.\ (\ref{eq:AtoI})].
For weak magnetic field, the spectral functions exhibit 
  similar behavior obtained in the SU(2) symmetric case,
\cite{meden_spectral,voit_spectral} except for the 
  non-zero weight seen at $-u_+q < \omega < -u_-q$.
In the SU(2) symmetric case, there is no weight at $-u_+q < \omega <
 -u_-q$
(see Fig.\ \ref{fig:spectral-SU2}).
For strong magnetic field, 
  the difference between $u_1$ and $u_2$ becomes large and 
 the spin-up and spin-down electrons are effectively decoupled,
  where $A_{R,1}$ ($A_{R,2}$) has stronger weight at
  $\omega\approx u_+q$ ($\omega\approx u_-q$).
We note that
the weight of $A_{R,2}(q,\omega)$ at $\omega \simeq u_- q$ becomes
   relatively large compared with that of 
$A_{R,1}(q,\omega)$ at $\omega \simeq u_+ q$, 
 reflecting  the large density of states for spin-down particle.
Especially, $A_{R,2}(q,\omega)$
  exhibits a similar behavior 
 obtained in the spinless case [Fig. 1(a) in Ref.\ \onlinecite{meden_spectral}].
We also note that,
 as seen from Fig.\ \ref{fig:contour}, 
  the two branch feature is prominent near Fermi energy ($\omega \simeq
 0$),
  however, one-particle feature is recovered in the  high-energy region.
As seen in Fig.\ \ref{fig:velocity}, the 
velocity $u_+$ takes a close value to $u_1$, while 
 $u_-$ is strongly renormalized from the bare velocity $u_2$.
For the spectral function $A_2(k,\omega)$, 
    a strong singularity can be seen at 
  $\omega = u_- q$, while 
  relatively large weight can be obtained at 
  $-|u_+ q| < \omega < |u_+q|$ for small $\omega$.

\begin{figure}[t]
\includegraphics[width=7cm]{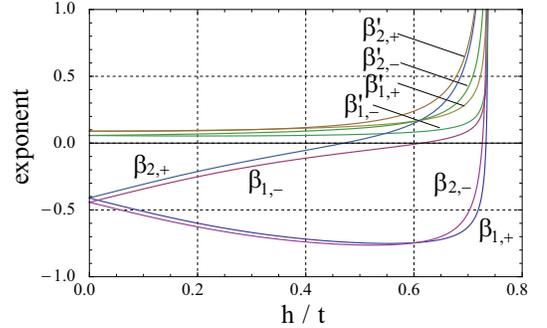}
\caption{
(Color online)
Exponents $\beta_{a,\pm}$ and $\beta'_{a,\pm}$
 as a function of the magnetic field
for $U/t=3$ and $\rho=1/2$.
}
\label{fig:exponent}
\end{figure}

\begin{figure}[t]
\includegraphics[width=5cm]{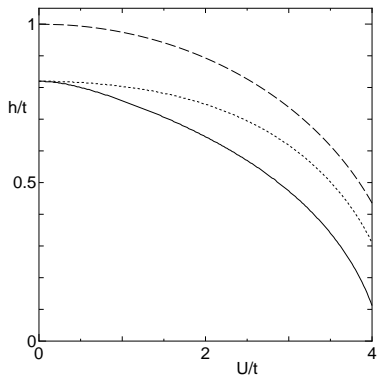}
\caption{
The $U$-dependence of 
the critical value for $h/t$ satisfying
  $\beta_{2,+}=0$ (solid line) and $\beta_{1,-}=0$ (dotted
line), for $\rho=1/2$.
 The dash line  denotes the  upper 
bound of 
 $h/t$ for the real $u_2$, i.e., corresponding to $u_{2c}$.
}

\label{fig:critical_value_h}
\end{figure}

By using Eqs.\ (\ref{eq:exponent_beta}) and (\ref{eq:mu}),
the $h$ dependence of 
the exponents $\beta$s are obtained (Fig.\ \ref{fig:exponent}).
As mentioned before, the spectral function  has two peaks 
 at $\omega = u_+q$ and $\omega = u_-q$ 
 for  $\beta_{2,+}<0$ and $\beta_{1,-}<0$, respectively. 
 while  each divergence is replaced by a cusp for  
 $\beta_{2,+}>0$ and $\beta_{1,-}>0$, respectively. 
Figure~\ref{fig:critical_value_h}
  shows such a critical value of $h/t$ as a function 
 of $U/t$, for $\beta_{2,+}=0$ (solid line) and $\beta_{1-}=0$ (dotted line).
The dash line corresponds to the upper bound of $h/t$ for the stable $u_-$. 
With increasing $U/t$, the critical value of $h/t$   
 for $\beta_{2,+}=0$ and $\beta_{1,-}=0$ 
decreases.

\section{Conclusion}
\label{sec:conclusion}

In the present paper, we have derived expressions of the fermion spectral
functions of a two-component Luttinger liquid at zero temperature in
terms of Appell hypergeometric functions. We have shown that in the
SU(2) symmetric case, these expressions reduce to the Gauss
hypergeometric functions. Our expressions allow the recovery of the
singularities derived in Refs.\ \onlinecite{meden_spectral,voit_spectral}
 but also describe the amplitude of the singularities as well as the
 behavior of the spectral function away from the singularities. The
 results of the present paper could be used to calculate zero
 temperature fermion
 spectral functions of various integrable
 models\cite{gaudin_fermions,yang_fermions,kawakami_tj,imambekov2006}
 using exact results on the TL-liquid exponents. In the case of
 three-component models,\cite{luscher2009} a more general version of
 the Feynman identity\cite{lebellac_qft} is applicable, and we expect
 that the zero temperature spectral function will also be expressible
 in terms of generalized hypergeometric functions of three variables.    
In the case of non-zero temperature, the real-space Green's function
is expressed as a product of powers of hyperbolic
sines\cite{nakamura_suzumura} and the Feynman
identity\cite{lebellac_qft} we used to derive our expressions does not
lead to a tractable expression. However, it is still possible to
discuss the qualitative changes to the spectral functions produced by
positive temperature. The first effect of being at finite temperature
is that the divergences of the spectral functions at $\omega=u_\nu q$
($\nu=\pm$) are cut-off by the thermal length which is of the
order of $u_-/T$. Therefore, the peaks of the spectral function
$A_{R,a}(q,\omega)$ will be replaced by
maxima of height $\sim T^{\beta_{a,\nu}}$ as $\omega \to u_\nu
q$. For $|\omega - u_\nu q|\gg T$, we expect that the integral giving
the spectral function is not strongly affected by the finite
temperature as the largest contribution comes from integration over
lengthscales smaller than the thermal length. Therefore, for the
lowest (non-zero) temperature scale, we expect that the spectral
function will be modified by the replacement of the factor $|\omega -
u_\nu q|^{\beta_{a,\nu}}$ by a scaling function $T^{\beta_{a,\nu}}
\mathcal{F}_\nu[(\omega -u_\nu q)/T]$ in
Eqs.~(\ref{eq:A2-final})--~(\ref{eq:A4-final}). The scaling function
will be such that $\mathcal{F}_\nu(0)$ is a constant, while
$\mathcal{F}_\nu(x) \sim x^{\beta_{a,\nu}}$ when $x \to \infty$. A
similar, but weaker effect should be seen on the cusps at
$\omega=-u_\pm q$. Again, the power laws will be replaced by scaling
functions that reproduce the infinite slope for $T=0$. Finally, we
expect that the gap $-u_- q < \omega < u_- q$ will start to fill.  
For higher temperatures, $T \gg |u_+-u_-|q$, we expect that the
difference between the peaks at $\omega=u_\pm q$ becomes blurred, and
only a single broad peak will be observed.

 \begin{acknowledgments}
{
E.O.\ acknowledges discussions with R.\ Citro and T.\ Ito.
M.T.\ thanks A.\ Furusaki for useful comments. 
} 
M.T.\ is supported in part by JSPS Institutional Program for 
Young Researcher Overseas Visits.
Y.S. is indebted to the Daiko foundation for the financial aid
 in the present work.
This work was financially supported in part
 by Grant-in-Aid for Special Coordination Funds for Promoting
Science and Technology (SCF), Scientific Research 
on Innovative Areas 20110002 
 from the Ministry of Education, Culture, Sports, Science and
 Technology, Japan. 
\end{acknowledgments}

\appendix

\section{Integrals}
\label{sec-app-integral}

\subsection{Fourier transforms} 
To obtain the spectral functions of the fermions, 
we need the Fourier transform:
\begin{eqnarray}
  \label{eq:transform}
J(q,\omega)=  
\int \frac{dx dt \, e^{-i(qx -\omega t)}}
{[\alpha + i(u_1 t-x)]^{\gamma_1} 
 [\alpha + i(u_2 t +x)]^{\gamma_2}},  
\end{eqnarray}
in the limit of $\alpha \to 0_+$. 
By the change of variables: 
\begin{eqnarray}
  \label{eq:linchvar}
  t=\frac{X_1+X_2}{u_1+u_2}, \quad
  x=\frac{u_1 X_2 - u_2 X_1}{u_1+u_2}, 
\end{eqnarray}
we find:
\begin{eqnarray}\label{eq:factorized-int}
  J(q,\omega)
&=&
\frac 1 {u_1+u_2} \int dX_1 \frac{e^{i \frac{(q u_2
      +\omega)}{u_1+u_2} X_1}}{(\alpha + i X_1)^{\gamma_1}}   
\nonumber \\ && {} \times
\int dX_2 \frac{e^{-i \frac{(q u_1
      -\omega)}{u_1+u_2} X_2}}{(\alpha + i X_2)^{\gamma_2}}.   
\end{eqnarray}
The two integrals in the product are obtained from the formula
(3.382.7) of Ref.~\onlinecite{gradshteyn80_tables}, giving the final
result:
\begin{eqnarray}
  \label{eq:transformed}
  J(q,\omega)&=&\frac{(2\pi)^2}{\Gamma(\gamma_1) \Gamma(\gamma_2)}
  \frac{(\omega+u_2 q)^{\gamma_1 -1}(\omega-u_1 q)^{\gamma_{{2}} -1}}{(u_1
    + u_2)^{\gamma_1+\gamma_2 -1}} \nonumber \\ & &\times  \Theta
  (\omega - u_1 q)  \Theta (\omega + u_2 q),  
\end{eqnarray}
where the limit $\alpha \to 0_+$ has been taken.

\subsection{Integrals expressible as hypergeometric functions}

The integral representation of the Gauss hypergeometric function 
 is
\begin{equation}
{}_2F_1(\alpha,\beta;\gamma;z)
=
\frac{\Gamma(\gamma)}{\Gamma(\beta)\Gamma(\gamma-\beta)} \int_0^1 dt \, 
\frac{t^{\beta-1} (1-t)^{\gamma-\beta-1}}{(1-tz)^{\alpha}},
\label{eq:appFdef}
\end{equation}
which satisfies 
[Eq. (15.3.3) in
Ref.\ \onlinecite{abramowitz_math_functions}]
\begin{equation}
{}_2F_1(\alpha,\beta;\gamma;z)
=(1-z)^{\gamma-\alpha-\beta} 
{}_2F_1(\gamma-\alpha,\gamma-\beta;\gamma;z).
\label{eq:appF}
\end{equation}

Typical integral formula for obtaining the spectral functions
(\ref{eq:master}) is given by
\begin{equation}
\bar I \equiv 
\int_0^1 dw \, w^{\alpha-1} (1-w)^{\beta-1}
 \frac{(a+bw)^{\gamma-1}}{(c - bw)^{\alpha+\beta+\gamma-1}} .
\end{equation}
By changing variable
$t = 1- c(1-w)/(c-bw)$,
we find
\begin{eqnarray}
\bar I 
&=&
\frac{(a+b)^{\beta+\gamma-1} }
{a^{\beta}(- b+c)^{\alpha+\beta+\gamma-1}}
\frac{\Gamma(\alpha)\Gamma(\beta)}{\Gamma(\alpha+\beta)}
\nonumber \\ && {} \times
 {}_2 F_1 \left(\alpha+\beta+\gamma-1, \, \beta; \alpha+\beta;
	     \frac{b(- a-c)}{a(- b+c)}\right),
\nonumber \\
\end{eqnarray}
where we have used Eqs.\ (\ref{eq:appFdef}) and
  (\ref{eq:appF}).

For the evaluation of Eq.\ (\ref{eq:A2A3}), 
we meet the integral:
\begin{equation}
I_1=
\int_0^1 dt \,  (1-t)^{\beta-1}  t^{\beta'-1} {}_2F_1(a,\beta+\beta';c;\lambda+\mu t).
\end{equation}
By changing the variable $t$ by
$t=\lambda\nu/(\lambda+\mu-\mu \nu)$, 
and by expanding
${}_2F_1\left(a,b;c;z \right)=\sum_{n=0}^\infty\frac{(a)_n(b)_n}{(c)_n}
\frac{z^n}{n!}$ where
 $(a)_n\equiv \Gamma(a+n)/\Gamma(a)$,
we find
\begin{widetext}
\begin{eqnarray}
I_1
&=&
\sum_{n=0}^\infty
 (\lambda+\mu)^\beta \lambda^{\beta'}
\frac{(a)_n(\beta+\beta')_n}{(c)_n} \frac{[\lambda(\lambda+\mu )]^n}{n!}
\int_0^1 d\nu \,
\frac{ (1-\nu)^{\beta-1} (\nu)^{\beta'-1} }{(\lambda+\mu-\mu \nu)^{\beta+\beta'+n}}
\nonumber \\
&=&
\frac{\Gamma(\beta)\Gamma(\beta')}{\Gamma(\beta+\beta')}
\left( \frac{\lambda}{\lambda+\mu}\right)^{\beta'}
\sum_{n=0}^\infty \frac{1}{n!}
\frac{(a)_n(\beta+\beta')_n}{(c)_n} 
{}_2F_1\left(\beta+\beta'+n,\beta';\beta+\beta';\frac{\mu}{\lambda+\mu} \right)
\lambda^n
\nonumber \\
&=&
\frac{\Gamma(\beta)\Gamma(\beta')}{\Gamma(\beta+\beta')}
\left(\frac{\lambda}{\lambda+\mu}\right)^{\beta'}
\sum_{n,m=0}^\infty 
\frac{(\beta+\beta')_{n+m}(a)_n (\beta')_m}{(c)_n (\beta'+\beta)_m} 
\frac{\lambda^n}{n!} \frac{(\frac{\mu}{\lambda+\mu})^m}{m!}
\nonumber \\
&=&
\frac{\Gamma(\beta)\Gamma(\beta')}{\Gamma(\beta+\beta')}
\left(\frac{\lambda}{\lambda+\mu}\right)^{\beta'}
F_2\left(\beta+\beta'; a,\beta';c,\beta+\beta'; \lambda,\frac{\mu}{\lambda+\mu}\right),
\end{eqnarray}
where $F_2$ is the Appell hypergeometric function.
By using the relation:
$F_2(A;B,B';C,A;x,y)
=
(1-y)^{-B'} F_1 \left(B; A-B',B';C; x , x/(1-y)\right)$,
we obtain: 
\begin{eqnarray}
I_1
&=&
B(\beta,\beta')
F_1 \left(a;\beta,\beta';c; \lambda , \lambda+\mu\right),
\label{eq:AppellF1}
\end{eqnarray}
where $F_1$ is the Appell hypergeometric function.
Similarly, 
the integral in  Eq.\ (\ref{eq:A1A4}) can be performed as: 
\begin{eqnarray}
I_2&=&
\int_0^1 dt \,  (1-t)^{\beta-1}  t^{\beta'-1} 
\left(1+\frac{\mu}{\lambda} t\right)^{c-1}
{}_2F_1(\beta+\beta'+c-1,b;c;\lambda+\mu t)
\nonumber  \\
&=&
\sum_{n=0}^\infty
\left(\frac{\lambda}{ \lambda+\mu } \right)^{\beta'}
\frac{(\beta+\beta'+c-1)_n(b)_n}{(c)_n} 
\frac{\lambda^n}{n!}
\int_0^1 d\nu \,
 (1-\nu)^{\beta-1} (\nu)^{\beta'-1} 
\left(1 - \frac{\mu \nu}{\lambda+\mu}\right)^{-(\beta+\beta'+c-1+n)}
\nonumber \\
&=&
\sum_{n=0}^\infty
\left(\frac{\lambda}{ \lambda+\mu } \right)^{\beta'}
\frac{(\beta+\beta'+c-1)_n(b)_n}{(c)_n} 
\frac{\lambda^n}{n!}
B(\beta,\beta')
{}_2F_1
\left(
\beta+\beta'+c-1+n, \beta'; \beta+\beta' ; \frac{\mu}{\lambda+\mu}
\right)
\nonumber \\
&=&
\left(\frac{\lambda}{ \lambda+\mu } \right)^{\beta'}
B(\beta,\beta')
\sum_{n,m=0}^\infty
\frac{\lambda^n}{n!}
\frac{(\frac{\mu}{\lambda+\mu})^n}{m!}
(\beta+\beta'+c-1)_{n+m} 
\frac{
(b)_n
(\beta')_m
}{
(c)_n
( \beta+\beta')_m
}
\nonumber  \\
&=&
\left(\frac{\lambda}{ \lambda+\mu } \right)^{\beta'}
B(\beta,\beta')
F_2 
\left(
\beta+\beta'+c-1; b, \beta' ; c,  \beta+\beta' ;
\lambda, \frac{\mu}{ \lambda+\mu }
\right).
\label{eq:AppellF2}
\end{eqnarray}

\end{widetext}

\section{Finite size bosonization and exponents} 

It is well known in conformal field theory that there is a relation
between the energy-momentum tensor and the Virasoro generators that
give the conformal weights (or dimensions) of the
operators\cite{difrancesco_book_conformal,cardy_conformal_book}. 
As a result, it is possible to relate the dimension of a given
operator to the energy of the state generated by acting on the ground
state of the Hamiltonian of a conformally invariant model with that
operator\cite{saleur_houches}. Multicomponent models are not in
general conformally invariant, but their critical properties can be
obtained by considering a semi-direct product of Virasoro
algebras\cite{frahm_confinv}. A more elementary approach, that we will
follow here, uses finite size
bosonization\cite{haldane_bosonisation,senechal_bosonization_revue}. 
In that approach, the fields $\phi_a$ and $\theta_a$ in
Eq.~(\ref{eq:bosonized}) admit the decomposition:
\begin{subequations}
\begin{eqnarray}
  \phi_a(x)&=&\phi_0^{(a)} -\frac{\pi n_a}{L} x + \frac 1 {\sqrt{L}}
  \sum_q \phi_a(q) e^{i q x}, \\ 
   \theta_a(x)&=&\theta_0^{(a)} -\frac{\pi J_a}{L} x + \frac 1 {\sqrt{L}}
  \sum_q \theta_a(q) e^{i q x}. 
\end{eqnarray}%
  \label{eq:finite-size-expansion}%
\end{subequations}
In Eqs.~(\ref{eq:finite-size-expansion}), we have the commutation
relations: $[\phi_0^{(a)},J_b]=-i \delta_{ab}$ and
$[\theta_0^{(a)},n_b]=-i \delta_{ab}$. In the Hamiltonian,
$\phi_0^{a}$ and $\theta_0^{(a)}$ do not appear as a result of the
derivations, but there is an extra term:
\begin{eqnarray}
  \delta H &=& \frac{\pi}{2L} \sum_{a,b} (M_{ab} J_a J_b + N_{ab} n_a
  n_b)
\nonumber \\ 
&=& \frac{\pi}{2L} ({}^t \bm J M \bm J
+
 {}^t \bm n N \bm n)   ,   
  \label{eq:zero-modes}
\end{eqnarray}
[$\bm J= {}^t(J_1,J_2)$ and $\bm n={}^t (n_1,n_2)$]
which vanishes when $L\to \infty$ and is called the zero mode
contribution. The ground state has $J_a=0$ and $n_a=0$. 
If we consider the momentum operator,
\begin{eqnarray}
  P=\sum_a \int dx \, \Pi_a \partial_x \phi_a, 
\end{eqnarray}
it also contains a zero mode contribution equal to:
\begin{eqnarray}
  \label{eq:momentum-zero-modes}
  \delta P = \frac \pi L \sum_a n_a J_a =\frac \pi L {}^t \bm n \bm J .
\end{eqnarray}

If we use the relations $M=Q(\Delta_2)^{1/2} {}^t Q$ and
$N=P(\Delta_2)^{1/2}  {}^tP$, we can rewrite:
\begin{eqnarray}
  \delta H &=& \frac{\pi}{4L} \left[ ({}^t \bm J Q + {}^t \bm n P)
    (\Delta_2)^{1/2} ({}^t Q \bm J + {}^t P \bm n) 
\right. \nonumber \\ && {} \left. 
+ ({}^t \bm J Q - {}^t \bm n P)
    (\Delta_2)^{1/2} ({}^t Q \bm J - {}^t P \bm n) \right] , \\
  \delta P &=&   \frac{\pi}{4L} \left[ ({}^t \bm J Q + {}^t \bm n P)
    ({}^t Q \bm J + {}^t P \bm n) 
\right. \nonumber \\ && {} \left. 
- ({}^t \bm J Q - {}^t \bm n P)
     ({}^t Q \bm J - {}^t P \bm n) \right] .
\end{eqnarray}

Now, if we act on the ground state with the operator $e^{i\sum_a (
  \eta_a \theta_a + \xi_a \phi_a)(x)}$, due to the presence of the
term $\sum_a (\eta_a \theta_0^{(a)}+\xi_a \phi_0^{(a)})$, the
resulting  state will belong to the subspace with
$n_a=-\eta_a,J_a=-\xi_a$. Its zero-mode contribution to the ground
state energy will be:
\begin{eqnarray}
   \delta H &=& \frac{\pi}{4L} \sum_\beta \left[ 
    u_\beta ({}^t \xi Q  + {}^t\eta P)_\beta^2 + 
    u_\beta  ({}^t \xi Q  - {}^t\eta P)_\beta^2 \right] ,
\nonumber \\ 
\\
  \delta P &=&   \frac{\pi}{4L} \left[ ({}^t \xi Q  + {}^t\eta
    P)_\beta^2  -({}^t \xi Q  - {}^t\eta P)_\beta^2  \right].
\end{eqnarray}

For an infinite system, a simple generalization of Eq.\ 
(\ref{eq:fermion-correlators}) shows that the correlation function take the form:
\begin{widetext}
\begin{eqnarray}
 \langle e^{i(\sum_a
  \eta_a \theta_a + \xi_a \phi_a)(x,t)} e^{-i(\sum_a
  \eta_a \theta_a + \xi_a \phi_a)(0,0)} \rangle 
= \prod_\beta
\left[ \frac \alpha {\alpha+ i(u_\beta t +x)} \right]^{({}^t\eta P +
  {}^t \xi Q)_\beta^2/4} 
\left[ \frac \alpha {\alpha+ i(u_\beta t -x)} \right]^{({}^t\eta P -
  {}^t \xi Q)_\beta^2/4}  .
\end{eqnarray}
\end{widetext}

So we see that the dimensions $({}^t\eta P \pm
  {}^t \xi Q)_\beta^2/4$  in the correlation functions also appear in
the zero mode contributions to the excited state energy and momentum of the finite
size system. This remark is the basis for the method of Frahm and
Korepin.\cite{frahm_confinv} Indeed, Eq.\ (8.2) of
Ref.~\onlinecite{frahm_confinv} is recovered with $P={}^t U
{}^t(Z^{-1})$ where:
\begin{equation}
  U=\left(
  \begin{array}{cc}
    1 & 1 \\ 0 & 1 
  \end{array}
\right), 
\end{equation}
since Frahm and Korepin have defined $N_c=N_\uparrow+N_\downarrow$ and
$N_s=N_\downarrow$, so that:
\begin{equation}
  \left(
    \begin{array}{c} 
      N_c \\ N_s 
    \end{array}
\right) = U  \left(
    \begin{array}{c} 
      N_\uparrow \\ N_\downarrow 
    \end{array}
\right).
\end{equation}
Similarly, Eq.\ (3.6) of Ref.~\onlinecite{frahm_confinv} is
recovered by taking $Q=U^{-1} Z$. Using these definitions, we can
recover Eqs.\ (3.12), (3.11) and (3.13) of
Ref.~\onlinecite{frahm_confinv}. { These relations have also been
derived in Ref.~\onlinecite{hikihara2005} by comparing the critical exponents
derived from the Bethe Ansatz with the ones derived from
bosonization.}


\begin{thebibliography}{89}
\expandafter\ifx\csname natexlab\endcsname\relax\def\natexlab#1{#1}\fi
\expandafter\ifx\csname bibnamefont\endcsname\relax
  \def\bibnamefont#1{#1}\fi
\expandafter\ifx\csname bibfnamefont\endcsname\relax
  \def\bibfnamefont#1{#1}\fi
\expandafter\ifx\csname citenamefont\endcsname\relax
  \def\citenamefont#1{#1}\fi
\expandafter\ifx\csname url\endcsname\relax
  \def\url#1{\texttt{#1}}\fi
\expandafter\ifx\csname urlprefix\endcsname\relax\def\urlprefix{URL }\fi
\providecommand{\bibinfo}[2]{#2}
\providecommand{\eprint}[2][]{\url{#2}}

\bibitem[{\citenamefont{Landau}(1957)}]{landau_fermiliquid_theory_static}
\bibinfo{author}{\bibfnamefont{L.~D.} \bibnamefont{Landau}},
  \bibinfo{journal}{Sov. Phys. JETP} \textbf{\bibinfo{volume}{3}},
  \bibinfo{pages}{920} (\bibinfo{year}{1957}).

\bibitem[{\citenamefont{Nozieres}(1961)}]{nozieres_book}
\bibinfo{author}{\bibfnamefont{P.}~\bibnamefont{Nozieres}},
  \emph{\bibinfo{title}{Theory of Interacting Fermi Systems}}
  (\bibinfo{publisher}{Benjamin}, \bibinfo{address}{New York},
  \bibinfo{year}{1961}).

\bibitem[{\citenamefont{Tomonaga}(1950)}]{tomonaga_model}
\bibinfo{author}{\bibfnamefont{S.}~\bibnamefont{Tomonaga}},
  \bibinfo{journal}{Prog. Theor. Phys.} \textbf{\bibinfo{volume}{5}},
  \bibinfo{pages}{544} (\bibinfo{year}{1950}).

\bibitem[{\citenamefont{Luttinger}(1963)}]{luttinger_model}
\bibinfo{author}{\bibfnamefont{J.~M.} \bibnamefont{Luttinger}},
  \bibinfo{journal}{J. Math. Phys.} \textbf{\bibinfo{volume}{4}},
  \bibinfo{pages}{1154} (\bibinfo{year}{1963}).

\bibitem[{\citenamefont{Schulz}(1995)}]{schulz_houches_revue}
\bibinfo{author}{\bibfnamefont{H.~J.} \bibnamefont{Schulz}}, in
  \emph{\bibinfo{booktitle}{Mesoscopic Quantum Physics, Les Houches LXI}},
  edited by \bibinfo{editor}{\bibfnamefont{E.}~\bibnamefont{Akkermans}},
  \bibinfo{editor}{\bibfnamefont{G.}~\bibnamefont{Montambaux}},
  \bibinfo{editor}{\bibfnamefont{J.~L.} \bibnamefont{Pichard}},
  \bibnamefont{and}
  \bibinfo{editor}{\bibfnamefont{J.}~\bibnamefont{Zinn-Justin}}
  (\bibinfo{publisher}{Elsevier}, \bibinfo{address}{Amsterdam},
  \bibinfo{year}{1995}), p. \bibinfo{pages}{533}.

\bibitem[{\citenamefont{Voit}(1995)}]{voit_bosonization_revue}
\bibinfo{author}{\bibfnamefont{J.}~\bibnamefont{Voit}}, \bibinfo{journal}{Rep.
  Prog. Phys.} \textbf{\bibinfo{volume}{58}}, \bibinfo{pages}{977}
  (\bibinfo{year}{1995}).

\bibitem[{\citenamefont{Varma et~al.}(2002)\citenamefont{Varma, Nussinov, and
  {van Saarloos}}}]{varma_nonfermiliquid_review}
\bibinfo{author}{\bibfnamefont{C.~M.} \bibnamefont{Varma}},
  \bibinfo{author}{\bibfnamefont{Z.}~\bibnamefont{Nussinov}}, \bibnamefont{and}
  \bibinfo{author}{\bibfnamefont{W.}~\bibnamefont{{van Saarloos}}},
  \bibinfo{journal}{Phys. Rep.} \textbf{\bibinfo{volume}{361}},
  \bibinfo{pages}{267} (\bibinfo{year}{2002}).

\bibitem[{\citenamefont{Suzumura}(1980)}]{suzumura1980}
\bibinfo{author}{\bibfnamefont{Y.}~\bibnamefont{Suzumura}},
  \bibinfo{journal}{Prog. Theor. Phys.} \textbf{\bibinfo{volume}{63}},
  \bibinfo{pages}{51} (\bibinfo{year}{1980}).

\bibitem[{\citenamefont{Claessen et~al.}(2002)\citenamefont{Claessen, Sing,
  Schwingenschl\"ogl, Blaha, Dressel, and Jacobsen}}]{Claessen2002}
\bibinfo{author}{\bibfnamefont{R.}~\bibnamefont{Claessen}},
  \bibinfo{author}{\bibfnamefont{M.}~\bibnamefont{Sing}},
  \bibinfo{author}{\bibfnamefont{U.}~\bibnamefont{Schwingenschl\"ogl}},
  \bibinfo{author}{\bibfnamefont{P.}~\bibnamefont{Blaha}},
  \bibinfo{author}{\bibfnamefont{M.}~\bibnamefont{Dressel}}, \bibnamefont{and}
  \bibinfo{author}{\bibfnamefont{C.~S.} \bibnamefont{Jacobsen}},
  \bibinfo{journal}{Phys. Rev. Lett.} \textbf{\bibinfo{volume}{88}},
  \bibinfo{pages}{096402} (\bibinfo{year}{2002}).

\bibitem[{\citenamefont{Ito et~al.}(2005)\citenamefont{Ito, Chainani, Haruna,
  Kanai, Yokoya, Shin, and Kato}}]{ito2005}
\bibinfo{author}{\bibfnamefont{T.}~\bibnamefont{Ito}},
  \bibinfo{author}{\bibfnamefont{A.}~\bibnamefont{Chainani}},
  \bibinfo{author}{\bibfnamefont{T.}~\bibnamefont{Haruna}},
  \bibinfo{author}{\bibfnamefont{K.}~\bibnamefont{Kanai}},
  \bibinfo{author}{\bibfnamefont{T.}~\bibnamefont{Yokoya}},
  \bibinfo{author}{\bibfnamefont{S.}~\bibnamefont{Shin}}, \bibnamefont{and}
  \bibinfo{author}{\bibfnamefont{R.}~\bibnamefont{Kato}},
  \bibinfo{journal}{Phys. Rev. Lett.} \textbf{\bibinfo{volume}{95}},
  \bibinfo{pages}{246402} (\bibinfo{year}{2005}).

\bibitem[{\citenamefont{J{\'e}rome}(1994)}]{jerome_organic_review}
\bibinfo{author}{\bibfnamefont{D.}~\bibnamefont{J{\'e}rome}}, in
  \emph{\bibinfo{booktitle}{Organic Conductors: fundamentals and
  applications}}, edited by \bibinfo{editor}{\bibfnamefont{J.-P.}
  \bibnamefont{Farges}} (\bibinfo{publisher}{Marcel Dekker},
  \bibinfo{address}{New York}, \bibinfo{year}{1994}), p. \bibinfo{pages}{405}.

\bibitem[{\citenamefont{Ijima}(1991)}]{ijima_nanotubes_synthesis}
\bibinfo{author}{\bibfnamefont{S.}~\bibnamefont{Ijima}},
  \bibinfo{journal}{Nature (London)} \textbf{\bibinfo{volume}{354}},
  \bibinfo{pages}{56} (\bibinfo{year}{1991}).

\bibitem[{\citenamefont{Dardel et~al.}(1991)\citenamefont{Dardel, Malterre,
  Grioni, Weibel, Baer, and L{\'e}vy}}]{dardel_photoemission_kmoo3}
\bibinfo{author}{\bibfnamefont{B.}~\bibnamefont{Dardel}},
  \bibinfo{author}{\bibfnamefont{D.}~\bibnamefont{Malterre}},
  \bibinfo{author}{\bibfnamefont{M.}~\bibnamefont{Grioni}},
  \bibinfo{author}{\bibfnamefont{P.}~\bibnamefont{Weibel}},
  \bibinfo{author}{\bibfnamefont{Y.}~\bibnamefont{Baer}}, \bibnamefont{and}
  \bibinfo{author}{\bibfnamefont{F.}~\bibnamefont{L{\'e}vy}},
  \bibinfo{journal}{Phys. Rev. Lett.} \textbf{\bibinfo{volume}{67}},
  \bibinfo{pages}{3144} (\bibinfo{year}{1991}).

\bibitem[{\citenamefont{Dardel et~al.}(1993)\citenamefont{Dardel, Malterre,
  Grioni, Weibel, Baer, Voit, and J{\'e}r{\^o}me}}]{dardel_photoemission}
\bibinfo{author}{\bibfnamefont{B.}~\bibnamefont{Dardel}},
  \bibinfo{author}{\bibfnamefont{D.}~\bibnamefont{Malterre}},
  \bibinfo{author}{\bibfnamefont{M.}~\bibnamefont{Grioni}},
  \bibinfo{author}{\bibfnamefont{P.}~\bibnamefont{Weibel}},
  \bibinfo{author}{\bibfnamefont{Y.}~\bibnamefont{Baer}},
  \bibinfo{author}{\bibfnamefont{J.}~\bibnamefont{Voit}}, \bibnamefont{and}
  \bibinfo{author}{\bibfnamefont{D.}~\bibnamefont{J{\'e}r{\^o}me}},
  \bibinfo{journal}{Europhys. Lett.} \textbf{\bibinfo{volume}{24}},
  \bibinfo{pages}{687} (\bibinfo{year}{1993}).

\bibitem[{\citenamefont{Gweon et~al.}(1996)\citenamefont{Gweon, Allen,
  Claessen, Clack, Poirier, Olson, Ellis, Zhang, Schneemeyer, Marcus
  et~al.}}]{gweon_bronzes_photoemission}
\bibinfo{author}{\bibfnamefont{G.~H.} \bibnamefont{Gweon}},
  \bibinfo{author}{\bibfnamefont{J.~W.} \bibnamefont{Allen}},
  \bibinfo{author}{\bibfnamefont{R.}~\bibnamefont{Claessen}},
  \bibinfo{author}{\bibfnamefont{J.~A.} \bibnamefont{Clack}},
  \bibinfo{author}{\bibfnamefont{D.~M.} \bibnamefont{Poirier}},
  \bibinfo{author}{\bibfnamefont{P.~J. B. C.~G.} \bibnamefont{Olson}},
  \bibinfo{author}{\bibfnamefont{W.~P.} \bibnamefont{Ellis}},
  \bibinfo{author}{\bibfnamefont{Y.}~\bibnamefont{Zhang}},
  \bibinfo{author}{\bibfnamefont{L.~F.} \bibnamefont{Schneemeyer}},
  \bibinfo{author}{\bibfnamefont{J.}~\bibnamefont{Marcus}},
  \bibnamefont{et~al.}, \bibinfo{journal}{J. Phys.: Condens. Matter}
  \textbf{\bibinfo{volume}{8}}, \bibinfo{pages}{9923} (\bibinfo{year}{1996}).

\bibitem[{\citenamefont{Mizokawa et~al.}(2002)\citenamefont{Mizokawa, Nakada,
  Kim, Shen, Yoshida, Fujimori, Horii, Yamada, Ikuta, and
  Mizutani}}]{mizokawa2002}
\bibinfo{author}{\bibfnamefont{T.}~\bibnamefont{Mizokawa}},
  \bibinfo{author}{\bibfnamefont{K.}~\bibnamefont{Nakada}},
  \bibinfo{author}{\bibfnamefont{C.}~\bibnamefont{Kim}},
  \bibinfo{author}{\bibfnamefont{Z.-X.} \bibnamefont{Shen}},
  \bibinfo{author}{\bibfnamefont{T.}~\bibnamefont{Yoshida}},
  \bibinfo{author}{\bibfnamefont{A.}~\bibnamefont{Fujimori}},
  \bibinfo{author}{\bibfnamefont{S.}~\bibnamefont{Horii}},
  \bibinfo{author}{\bibfnamefont{Y.}~\bibnamefont{Yamada}},
  \bibinfo{author}{\bibfnamefont{H.}~\bibnamefont{Ikuta}}, \bibnamefont{and}
  \bibinfo{author}{\bibfnamefont{U.}~\bibnamefont{Mizutani}},
  \bibinfo{journal}{Phys. Rev. B} \textbf{\bibinfo{volume}{65}},
  \bibinfo{pages}{193101} (\bibinfo{year}{2002}).

\bibitem[{\citenamefont{Ishii et~al.}(2003)\citenamefont{Ishii, Kataura,
  Shiozawa, Yoshioka, Otsubo, Takayama, Miyahara, Suzuki, Achiba, Nakatake
  et~al.}}]{ishii03_nanotube_pes}
\bibinfo{author}{\bibfnamefont{H.}~\bibnamefont{Ishii}},
  \bibinfo{author}{\bibfnamefont{H.}~\bibnamefont{Kataura}},
  \bibinfo{author}{\bibfnamefont{H.}~\bibnamefont{Shiozawa}},
  \bibinfo{author}{\bibfnamefont{H.}~\bibnamefont{Yoshioka}},
  \bibinfo{author}{\bibfnamefont{H.}~\bibnamefont{Otsubo}},
  \bibinfo{author}{\bibfnamefont{Y.}~\bibnamefont{Takayama}},
  \bibinfo{author}{\bibfnamefont{T.}~\bibnamefont{Miyahara}},
  \bibinfo{author}{\bibfnamefont{S.}~\bibnamefont{Suzuki}},
  \bibinfo{author}{\bibfnamefont{Y.}~\bibnamefont{Achiba}},
  \bibinfo{author}{\bibfnamefont{M.}~\bibnamefont{Nakatake}},
  \bibnamefont{et~al.}, \bibinfo{journal}{Nature (London)}
  \textbf{\bibinfo{volume}{426}}, \bibinfo{pages}{540} (\bibinfo{year}{2003}).

\bibitem[{\citenamefont{Denlinger et~al.}(1999)\citenamefont{Denlinger, Gweon,
  Allen, Olson, Marcus, Schlenker, and Hsu}}]{denlinger_arpes_LiMoO}
\bibinfo{author}{\bibfnamefont{J.~D.} \bibnamefont{Denlinger}},
  \bibinfo{author}{\bibfnamefont{G.-H.} \bibnamefont{Gweon}},
  \bibinfo{author}{\bibfnamefont{J.~W.} \bibnamefont{Allen}},
  \bibinfo{author}{\bibfnamefont{C.~G.} \bibnamefont{Olson}},
  \bibinfo{author}{\bibfnamefont{J.}~\bibnamefont{Marcus}},
  \bibinfo{author}{\bibfnamefont{C.}~\bibnamefont{Schlenker}},
  \bibnamefont{and} \bibinfo{author}{\bibfnamefont{L.-S.} \bibnamefont{Hsu}},
  \bibinfo{journal}{Phys. Rev. Lett.} \textbf{\bibinfo{volume}{82}},
  \bibinfo{pages}{2540} (\bibinfo{year}{1999}).

\bibitem[{\citenamefont{{Sing} et~al.}(2003)\citenamefont{{Sing},
  {Schwingenschl{\"o}gl}, {Claessen}, {Blaha}, {Carmelo}, {Martelo},
  {Sacramento}, {Dressel}, and {Jacobsen}}}]{sing2003}
\bibinfo{author}{\bibfnamefont{M.}~\bibnamefont{{Sing}}},
  \bibinfo{author}{\bibfnamefont{U.}~\bibnamefont{{Schwingenschl{\"o}gl}}},
  \bibinfo{author}{\bibfnamefont{R.}~\bibnamefont{{Claessen}}},
  \bibinfo{author}{\bibfnamefont{P.}~\bibnamefont{{Blaha}}},
  \bibinfo{author}{\bibfnamefont{J.~M.} \bibnamefont{{Carmelo}}},
  \bibinfo{author}{\bibfnamefont{L.~M.} \bibnamefont{{Martelo}}},
  \bibinfo{author}{\bibfnamefont{P.~D.} \bibnamefont{{Sacramento}}},
  \bibinfo{author}{\bibfnamefont{M.}~\bibnamefont{{Dressel}}},
  \bibnamefont{and} \bibinfo{author}{\bibfnamefont{C.~S.}
  \bibnamefont{{Jacobsen}}}, \bibinfo{journal}{Phys. Rev. B}
  \textbf{\bibinfo{volume}{68}}, \bibinfo{pages}{125111}
  (\bibinfo{year}{2003}).

\bibitem[{\citenamefont{Gweon et~al.}(2003)\citenamefont{Gweon, Mo, Allen, He,
  Jin, Mandrus, and H{\"o}chst}}]{gweon03_limo6o17_arpes}
\bibinfo{author}{\bibfnamefont{G.-H.} \bibnamefont{Gweon}},
  \bibinfo{author}{\bibfnamefont{S.-K.} \bibnamefont{Mo}},
  \bibinfo{author}{\bibfnamefont{J.~W.} \bibnamefont{Allen}},
  \bibinfo{author}{\bibfnamefont{J.}~\bibnamefont{He}},
  \bibinfo{author}{\bibfnamefont{R.}~\bibnamefont{Jin}},
  \bibinfo{author}{\bibfnamefont{D.}~\bibnamefont{Mandrus}}, \bibnamefont{and}
  \bibinfo{author}{\bibfnamefont{H.}~\bibnamefont{H{\"o}chst}},
  \emph{\bibinfo{title}{{Luttinger liquid ARPES spectra from samples of
  Li$_{0.9}$Mo$_6$O$_{17}$ grown by the temperature gradient flux technique}}},
  \bibinfo{howpublished}{cond-mat/0403008} (\bibinfo{year}{2003}).

\bibitem[{\citenamefont{Grioni}(2004)}]{grioni2004}
\bibinfo{author}{\bibfnamefont{M.}~\bibnamefont{Grioni}}, in
  \emph{\bibinfo{booktitle}{Strong interactions in low dimensions}}, edited by
  \bibinfo{editor}{\bibfnamefont{D.}~\bibnamefont{Baeriswyl}} \bibnamefont{and}
  \bibinfo{editor}{\bibfnamefont{L.}~\bibnamefont{Degiorgi}}
  (\bibinfo{publisher}{Springer}, \bibinfo{address}{Heidelberg, Germany},
  \bibinfo{year}{2004}), vol.~\bibinfo{volume}{25} of
  \emph{\bibinfo{series}{Physics and Chemistry of Materials with
  Low-Dimensional Structures}}, chap.~\bibinfo{chapter}{5}, p.
  \bibinfo{pages}{137}.

\bibitem[{\citenamefont{Vescoli et~al.}(2000)\citenamefont{Vescoli, Zwick,
  Henderson, DeGiorgi, Grioni, Gruner, and
  Montgomery}}]{vescoli_photoemission_tmtsf}
\bibinfo{author}{\bibfnamefont{V.}~\bibnamefont{Vescoli}},
  \bibinfo{author}{\bibfnamefont{F.}~\bibnamefont{Zwick}},
  \bibinfo{author}{\bibfnamefont{W.}~\bibnamefont{Henderson}},
  \bibinfo{author}{\bibfnamefont{L.}~\bibnamefont{DeGiorgi}},
  \bibinfo{author}{\bibfnamefont{M.}~\bibnamefont{Grioni}},
  \bibinfo{author}{\bibfnamefont{G.}~\bibnamefont{Gruner}}, \bibnamefont{and}
  \bibinfo{author}{\bibfnamefont{L.~K.} \bibnamefont{Montgomery}},
  \bibinfo{journal}{Eur. Phys. J. B} \textbf{\bibinfo{volume}{13}},
  \bibinfo{pages}{503} (\bibinfo{year}{2000}).

\bibitem[{\citenamefont{Himpsel et~al.}(2001)\citenamefont{Himpsel, Altmann,
  Bennewitz, Crain, Kirakosian, Lin, and McChesney}}]{himpsel2001}
\bibinfo{author}{\bibfnamefont{F.~J.} \bibnamefont{Himpsel}},
  \bibinfo{author}{\bibfnamefont{K.~N.} \bibnamefont{Altmann}},
  \bibinfo{author}{\bibfnamefont{R.}~\bibnamefont{Bennewitz}},
  \bibinfo{author}{\bibfnamefont{J.~N.} \bibnamefont{Crain}},
  \bibinfo{author}{\bibfnamefont{A.}~\bibnamefont{Kirakosian}},
  \bibinfo{author}{\bibfnamefont{J.-L.} \bibnamefont{Lin}}, \bibnamefont{and}
  \bibinfo{author}{\bibfnamefont{J.~L.} \bibnamefont{McChesney}},
  \bibinfo{journal}{J. Phys.: Condens. Matter} \textbf{\bibinfo{volume}{13}},
  \bibinfo{pages}{11097} (\bibinfo{year}{2001}).

\bibitem[{\citenamefont{Oncel}(2008)}]{oncel2008}
\bibinfo{author}{\bibfnamefont{N.}~\bibnamefont{Oncel}}, \bibinfo{journal}{J.
  Phys.: Condens. Matter} \textbf{\bibinfo{volume}{20}},
  \bibinfo{pages}{393001} (\bibinfo{year}{2008}).

\bibitem[{\citenamefont{Losio et~al.}(2001)\citenamefont{Losio, Altmann,
  Kirakosian, Lin, Petrovykh, and Himpsel}}]{Losio2001}
\bibinfo{author}{\bibfnamefont{R.}~\bibnamefont{Losio}},
  \bibinfo{author}{\bibfnamefont{K.~N.} \bibnamefont{Altmann}},
  \bibinfo{author}{\bibfnamefont{A.}~\bibnamefont{Kirakosian}},
  \bibinfo{author}{\bibfnamefont{J.-L.} \bibnamefont{Lin}},
  \bibinfo{author}{\bibfnamefont{D.~Y.} \bibnamefont{Petrovykh}},
  \bibnamefont{and} \bibinfo{author}{\bibfnamefont{F.~J.}
  \bibnamefont{Himpsel}}, \bibinfo{journal}{Phys. Rev. Lett.}
  \textbf{\bibinfo{volume}{86}}, \bibinfo{pages}{4632} (\bibinfo{year}{2001}).

\bibitem[{\citenamefont{Crain et~al.}(2004)\citenamefont{Crain, McChesney,
  Zheng, Gallagher, Snijders, Bissen, Gundelach, Erwin, and
  Himpsel}}]{Crain2004}
\bibinfo{author}{\bibfnamefont{J.~N.} \bibnamefont{Crain}},
  \bibinfo{author}{\bibfnamefont{J.~L.} \bibnamefont{McChesney}},
  \bibinfo{author}{\bibfnamefont{F.}~\bibnamefont{Zheng}},
  \bibinfo{author}{\bibfnamefont{M.~C.} \bibnamefont{Gallagher}},
  \bibinfo{author}{\bibfnamefont{P.~C.} \bibnamefont{Snijders}},
  \bibinfo{author}{\bibfnamefont{M.}~\bibnamefont{Bissen}},
  \bibinfo{author}{\bibfnamefont{C.}~\bibnamefont{Gundelach}},
  \bibinfo{author}{\bibfnamefont{S.~C.} \bibnamefont{Erwin}}, \bibnamefont{and}
  \bibinfo{author}{\bibfnamefont{F.~J.} \bibnamefont{Himpsel}},
  \bibinfo{journal}{Phys. Rev. B} \textbf{\bibinfo{volume}{69}},
  \bibinfo{pages}{125401} (\bibinfo{year}{2004}).

\bibitem[{\citenamefont{Okuda et~al.}(2010)\citenamefont{Okuda, Miyamaoto,
  Takeichi, Miyahara, Ogawa, Harasawa, Kimura, Matsuda, Kakizaki, Shishidou
  et~al.}}]{Okuda2010}
\bibinfo{author}{\bibfnamefont{T.}~\bibnamefont{Okuda}},
  \bibinfo{author}{\bibfnamefont{K.}~\bibnamefont{Miyamaoto}},
  \bibinfo{author}{\bibfnamefont{Y.}~\bibnamefont{Takeichi}},
  \bibinfo{author}{\bibfnamefont{H.}~\bibnamefont{Miyahara}},
  \bibinfo{author}{\bibfnamefont{M.}~\bibnamefont{Ogawa}},
  \bibinfo{author}{\bibfnamefont{A.}~\bibnamefont{Harasawa}},
  \bibinfo{author}{\bibfnamefont{A.}~\bibnamefont{Kimura}},
  \bibinfo{author}{\bibfnamefont{I.}~\bibnamefont{Matsuda}},
  \bibinfo{author}{\bibfnamefont{A.}~\bibnamefont{Kakizaki}},
  \bibinfo{author}{\bibfnamefont{T.}~\bibnamefont{Shishidou}},
  \bibnamefont{et~al.}, \bibinfo{journal}{Phys. Rev. B}
  \textbf{\bibinfo{volume}{82}}, \bibinfo{pages}{161410}
  (\bibinfo{year}{2010}).

\bibitem[{\citenamefont{Altland et~al.}(1999)\citenamefont{Altland, Barnes,
  Hekking, and Schofield}}]{altland1999}
\bibinfo{author}{\bibfnamefont{A.}~\bibnamefont{Altland}},
  \bibinfo{author}{\bibfnamefont{C.~H.~W.} \bibnamefont{Barnes}},
  \bibinfo{author}{\bibfnamefont{F.~W.~J.} \bibnamefont{Hekking}},
  \bibnamefont{and} \bibinfo{author}{\bibfnamefont{A.~J.}
  \bibnamefont{Schofield}}, \bibinfo{journal}{Phys. Rev. Lett.}
  \textbf{\bibinfo{volume}{83}}, \bibinfo{pages}{1203} (\bibinfo{year}{1999}).

\bibitem[{\citenamefont{Grigera et~al.}(2004)\citenamefont{Grigera, Schofield,
  Rabello, and Si}}]{grigera2004}
\bibinfo{author}{\bibfnamefont{S.~A.} \bibnamefont{Grigera}},
  \bibinfo{author}{\bibfnamefont{A.~J.} \bibnamefont{Schofield}},
  \bibinfo{author}{\bibfnamefont{S.}~\bibnamefont{Rabello}}, \bibnamefont{and}
  \bibinfo{author}{\bibfnamefont{Q.}~\bibnamefont{Si}}, \bibinfo{journal}{Phys.
  Rev. B} \textbf{\bibinfo{volume}{69}}, \bibinfo{pages}{245109}
  (\bibinfo{year}{2004}).

\bibitem[{\citenamefont{Carpentier et~al.}(2002)\citenamefont{Carpentier, Peca,
  and Balents}}]{peca_tunneling_quantumwire}
\bibinfo{author}{\bibfnamefont{D.}~\bibnamefont{Carpentier}},
  \bibinfo{author}{\bibfnamefont{C.}~\bibnamefont{Peca}}, \bibnamefont{and}
  \bibinfo{author}{\bibfnamefont{L.}~\bibnamefont{Balents}},
  \bibinfo{journal}{Phys. Rev. B} \textbf{\bibinfo{volume}{66}},
  \bibinfo{pages}{153304} (\bibinfo{year}{2002}).

\bibitem[{\citenamefont{Auslaender et~al.}(2005)\citenamefont{Auslaender,
  Steinberg, Yacoby, Tserkovnyak, Halperin, Baldwin, Pfeiffer, and
  West}}]{auslaender2005}
\bibinfo{author}{\bibfnamefont{O.~M.} \bibnamefont{Auslaender}},
  \bibinfo{author}{\bibfnamefont{H.}~\bibnamefont{Steinberg}},
  \bibinfo{author}{\bibfnamefont{A.}~\bibnamefont{Yacoby}},
  \bibinfo{author}{\bibfnamefont{Y.}~\bibnamefont{Tserkovnyak}},
  \bibinfo{author}{\bibfnamefont{B.~I.} \bibnamefont{Halperin}},
  \bibinfo{author}{\bibfnamefont{K.~W.} \bibnamefont{Baldwin}},
  \bibinfo{author}{\bibfnamefont{L.~N.} \bibnamefont{Pfeiffer}},
  \bibnamefont{and} \bibinfo{author}{\bibfnamefont{K.~W.} \bibnamefont{West}},
  \bibinfo{journal}{Science} \textbf{\bibinfo{volume}{308}},
  \bibinfo{pages}{88} (\bibinfo{year}{2005}).

\bibitem[{\citenamefont{Jompol et~al.}(2009)\citenamefont{Jompol, Ford,
  Griffiths, Farrer, Jones, Anderson, Ritchie, Silk, and
  Schofield}}]{jompol2009}
\bibinfo{author}{\bibfnamefont{Y.}~\bibnamefont{Jompol}},
  \bibinfo{author}{\bibfnamefont{C.~J.~B.} \bibnamefont{Ford}},
  \bibinfo{author}{\bibfnamefont{J.~P.} \bibnamefont{Griffiths}},
  \bibinfo{author}{\bibfnamefont{I.}~\bibnamefont{Farrer}},
  \bibinfo{author}{\bibfnamefont{G.~A.~C.} \bibnamefont{Jones}},
  \bibinfo{author}{\bibfnamefont{D.}~\bibnamefont{Anderson}},
  \bibinfo{author}{\bibfnamefont{D.~A.} \bibnamefont{Ritchie}},
  \bibinfo{author}{\bibfnamefont{T.~W.} \bibnamefont{Silk}}, \bibnamefont{and}
  \bibinfo{author}{\bibfnamefont{A.~J.} \bibnamefont{Schofield}},
  \bibinfo{journal}{Science} \textbf{\bibinfo{volume}{325}},
  \bibinfo{pages}{597} (\bibinfo{year}{2009}).

\bibitem[{\citenamefont{{Recati}
  et~al.}(2003{\natexlab{a}})\citenamefont{{Recati}, {Fedichev}, {Zwerger}, and
  {Zoller}}}]{recati2003}
\bibinfo{author}{\bibfnamefont{A.}~\bibnamefont{{Recati}}},
  \bibinfo{author}{\bibfnamefont{P.~O.} \bibnamefont{{Fedichev}}},
  \bibinfo{author}{\bibfnamefont{W.}~\bibnamefont{{Zwerger}}},
  \bibnamefont{and} \bibinfo{author}{\bibfnamefont{P.}~\bibnamefont{{Zoller}}},
  \bibinfo{journal}{Phys. Rev. Lett.} \textbf{\bibinfo{volume}{90}},
  \bibinfo{pages}{020401} (\bibinfo{year}{2003}{\natexlab{a}}).

\bibitem[{\citenamefont{{Recati}
  et~al.}(2003{\natexlab{b}})\citenamefont{{Recati}, {Fedichev}, {Zwerger}, and
  {Zoller}}}]{recati2003a}
\bibinfo{author}{\bibfnamefont{A.}~\bibnamefont{{Recati}}},
  \bibinfo{author}{\bibfnamefont{P.~O.} \bibnamefont{{Fedichev}}},
  \bibinfo{author}{\bibfnamefont{W.}~\bibnamefont{{Zwerger}}},
  \bibnamefont{and} \bibinfo{author}{\bibfnamefont{P.}~\bibnamefont{{Zoller}}},
  \bibinfo{journal}{J. Phys. B} \textbf{\bibinfo{volume}{5}},
  \bibinfo{pages}{55} (\bibinfo{year}{2003}{\natexlab{b}}).

\bibitem[{\citenamefont{Kollath and {Schollw\"ock}}(2006)}]{kollath2006}
\bibinfo{author}{\bibfnamefont{C.}~\bibnamefont{Kollath}} \bibnamefont{and}
  \bibinfo{author}{\bibfnamefont{U.}~\bibnamefont{{Schollw\"ock}}},
  \bibinfo{journal}{New J. Phys.} \textbf{\bibinfo{volume}{8}},
  \bibinfo{pages}{220} (\bibinfo{year}{2006}).

\bibitem[{\citenamefont{{Paredes {\it et al.}}}(2004)}]{paredes_tonks_optical}
\bibinfo{author}{\bibfnamefont{B.}~\bibnamefont{{Paredes {\it et al.}}}},
  \bibinfo{journal}{Nature (London)} \textbf{\bibinfo{volume}{429}},
  \bibinfo{pages}{277} (\bibinfo{year}{2004}).

\bibitem[{\citenamefont{Kinoshita et~al.}(2004)\citenamefont{Kinoshita, Wenger,
  and Weiss}}]{kinoshita_tonks_continuous}
\bibinfo{author}{\bibfnamefont{T.}~\bibnamefont{Kinoshita}},
  \bibinfo{author}{\bibfnamefont{T.}~\bibnamefont{Wenger}}, \bibnamefont{and}
  \bibinfo{author}{\bibfnamefont{D.~S.} \bibnamefont{Weiss}},
  \bibinfo{journal}{Science} \textbf{\bibinfo{volume}{305}},
  \bibinfo{pages}{1125} (\bibinfo{year}{2004}).

\bibitem[{\citenamefont{K\"ohl et~al.}(2004)\citenamefont{K\"ohl, St\"oferle,
  Moritz, Schori, and Esslinger}}]{koehl_nofk}
\bibinfo{author}{\bibfnamefont{M.}~\bibnamefont{K\"ohl}},
  \bibinfo{author}{\bibfnamefont{T.}~\bibnamefont{St\"oferle}},
  \bibinfo{author}{\bibfnamefont{H.}~\bibnamefont{Moritz}},
  \bibinfo{author}{\bibfnamefont{C.}~\bibnamefont{Schori}}, \bibnamefont{and}
  \bibinfo{author}{\bibfnamefont{T.}~\bibnamefont{Esslinger}},
  \bibinfo{journal}{Appl. Phys. B} \textbf{\bibinfo{volume}{79}},
  \bibinfo{pages}{1009} (\bibinfo{year}{2004}).

\bibitem[{\citenamefont{{van Amerongen} et~al.}(2008)\citenamefont{{van
  Amerongen}, {van Es}, {Wicke}, {Kheruntsyan}, and {van
  Druten}}}]{amerongen2008}
\bibinfo{author}{\bibfnamefont{A.~H.} \bibnamefont{{van Amerongen}}},
  \bibinfo{author}{\bibfnamefont{J.~J.~P.} \bibnamefont{{van Es}}},
  \bibinfo{author}{\bibfnamefont{P.}~\bibnamefont{{Wicke}}},
  \bibinfo{author}{\bibfnamefont{K.~V.} \bibnamefont{{Kheruntsyan}}},
  \bibnamefont{and} \bibinfo{author}{\bibfnamefont{N.~J.} \bibnamefont{{van
  Druten}}}, \bibinfo{journal}{Phys. Rev. Lett.}
  \textbf{\bibinfo{volume}{100}}, \bibinfo{pages}{090402}
  (\bibinfo{year}{2008}), \eprint{arXiv:0709.1899}.

\bibitem[{\citenamefont{{Bouchoule} et~al.}(2009)\citenamefont{{Bouchoule},
  {Van Druten}, and {Westbrook}}}]{bouchoule2009}
\bibinfo{author}{\bibfnamefont{I.}~\bibnamefont{{Bouchoule}}},
  \bibinfo{author}{\bibfnamefont{N.~J.} \bibnamefont{{Van Druten}}},
  \bibnamefont{and} \bibinfo{author}{\bibfnamefont{C.~I.}
  \bibnamefont{{Westbrook}}}, \emph{\bibinfo{title}{{Atom chips and
  one-dimensional Bose gases}}} (\bibinfo{year}{2009}),
  \eprint{arXiv:0901.3303}.

\bibitem[{\citenamefont{{G{\"u}nter} et~al.}(2006)\citenamefont{{G{\"u}nter},
  {St{\"o}ferle}, {Moritz}, {K{\"o}hl}, and {Esslinger}}}]{guenter2006}
\bibinfo{author}{\bibfnamefont{K.}~\bibnamefont{{G{\"u}nter}}},
  \bibinfo{author}{\bibfnamefont{T.}~\bibnamefont{{St{\"o}ferle}}},
  \bibinfo{author}{\bibfnamefont{H.}~\bibnamefont{{Moritz}}},
  \bibinfo{author}{\bibfnamefont{M.}~\bibnamefont{{K{\"o}hl}}},
  \bibnamefont{and}
  \bibinfo{author}{\bibfnamefont{T.}~\bibnamefont{{Esslinger}}},
  \bibinfo{journal}{Phys. Rev. Lett.} \textbf{\bibinfo{volume}{96}},
  \bibinfo{pages}{180402} (\bibinfo{year}{2006}).

\bibitem[{\citenamefont{{Ospelkaus} et~al.}(2006)\citenamefont{{Ospelkaus},
  {Ospelkaus}, {Wille}, {Succo}, {Ernst}, {Sengstock}, and
  {Bongs}}}]{ospelkaus2006}
\bibinfo{author}{\bibfnamefont{S.}~\bibnamefont{{Ospelkaus}}},
  \bibinfo{author}{\bibfnamefont{C.}~\bibnamefont{{Ospelkaus}}},
  \bibinfo{author}{\bibfnamefont{O.}~\bibnamefont{{Wille}}},
  \bibinfo{author}{\bibfnamefont{M.}~\bibnamefont{{Succo}}},
  \bibinfo{author}{\bibfnamefont{P.}~\bibnamefont{{Ernst}}},
  \bibinfo{author}{\bibfnamefont{K.}~\bibnamefont{{Sengstock}}},
  \bibnamefont{and} \bibinfo{author}{\bibfnamefont{K.}~\bibnamefont{{Bongs}}},
  \bibinfo{journal}{Physical Review Letters} \textbf{\bibinfo{volume}{96}},
  \bibinfo{pages}{180403} (\bibinfo{year}{2006}),
  \eprint{arXiv:cond-mat/0604179}.

\bibitem[{\citenamefont{Wille et~al.}(2008)\citenamefont{Wille, Spiegelhalder,
  Kerner, Naik, Trenkwalder, Hendl, Schreck, Grimm, Tiecke, Walraven
  et~al.}}]{wille2008}
\bibinfo{author}{\bibfnamefont{E.}~\bibnamefont{Wille}},
  \bibinfo{author}{\bibfnamefont{F.~M.} \bibnamefont{Spiegelhalder}},
  \bibinfo{author}{\bibfnamefont{G.}~\bibnamefont{Kerner}},
  \bibinfo{author}{\bibfnamefont{D.}~\bibnamefont{Naik}},
  \bibinfo{author}{\bibfnamefont{A.}~\bibnamefont{Trenkwalder}},
  \bibinfo{author}{\bibfnamefont{G.}~\bibnamefont{Hendl}},
  \bibinfo{author}{\bibfnamefont{F.}~\bibnamefont{Schreck}},
  \bibinfo{author}{\bibfnamefont{R.}~\bibnamefont{Grimm}},
  \bibinfo{author}{\bibfnamefont{T.~G.} \bibnamefont{Tiecke}},
  \bibinfo{author}{\bibfnamefont{J.~T.~M.} \bibnamefont{Walraven}},
  \bibnamefont{et~al.}, \bibinfo{journal}{Phys. Rev. Lett.}
  \textbf{\bibinfo{volume}{100}}, \bibinfo{pages}{053201}
  (\bibinfo{year}{2008}).

\bibitem[{\citenamefont{Taglieber et~al.}(2008)\citenamefont{Taglieber, Voigt,
  Aoki, H\"ansch, and Dieckmann}}]{taglieber2008}
\bibinfo{author}{\bibfnamefont{M.}~\bibnamefont{Taglieber}},
  \bibinfo{author}{\bibfnamefont{A.-C.} \bibnamefont{Voigt}},
  \bibinfo{author}{\bibfnamefont{T.}~\bibnamefont{Aoki}},
  \bibinfo{author}{\bibfnamefont{T.~W.} \bibnamefont{H\"ansch}},
  \bibnamefont{and}
  \bibinfo{author}{\bibfnamefont{K.}~\bibnamefont{Dieckmann}},
  \bibinfo{journal}{Phys. Rev. Lett.} \textbf{\bibinfo{volume}{100}},
  \bibinfo{pages}{010401} (\bibinfo{year}{2008}).

\bibitem[{\citenamefont{DeMarco and Jin}(1999)}]{DeMarco1999}
\bibinfo{author}{\bibfnamefont{B.}~\bibnamefont{DeMarco}} \bibnamefont{and}
  \bibinfo{author}{\bibfnamefont{D.~S.} \bibnamefont{Jin}},
  \bibinfo{journal}{Science} \textbf{\bibinfo{volume}{285}},
  \bibinfo{pages}{1703} (\bibinfo{year}{1999}).

\bibitem[{\citenamefont{{Liao} et~al.}(2010)\citenamefont{{Liao}, {Rittner},
  {Paprotta}, {Li}, {Partridge}, {Hulet}, {Baur}, and {Mueller}}}]{liao2010}
\bibinfo{author}{\bibfnamefont{Y.}~\bibnamefont{{Liao}}},
  \bibinfo{author}{\bibfnamefont{A.~S.~C.} \bibnamefont{{Rittner}}},
  \bibinfo{author}{\bibfnamefont{T.}~\bibnamefont{{Paprotta}}},
  \bibinfo{author}{\bibfnamefont{W.}~\bibnamefont{{Li}}},
  \bibinfo{author}{\bibfnamefont{G.~B.} \bibnamefont{{Partridge}}},
  \bibinfo{author}{\bibfnamefont{R.~G.} \bibnamefont{{Hulet}}},
  \bibinfo{author}{\bibfnamefont{S.~K.} \bibnamefont{{Baur}}},
  \bibnamefont{and} \bibinfo{author}{\bibfnamefont{E.~J.}
  \bibnamefont{{Mueller}}}, \bibinfo{journal}{Nature (London)}
  \textbf{\bibinfo{volume}{467}}, \bibinfo{pages}{567} (\bibinfo{year}{2010}),
  \eprint{0912.0092}.

\bibitem[{\citenamefont{Stewart et~al.}(2008)\citenamefont{Stewart, Gaebler,
  and Jin}}]{stewart2008}
\bibinfo{author}{\bibfnamefont{J.~T.} \bibnamefont{Stewart}},
  \bibinfo{author}{\bibfnamefont{J.~P.} \bibnamefont{Gaebler}},
  \bibnamefont{and} \bibinfo{author}{\bibfnamefont{D.~S.} \bibnamefont{Jin}},
  \bibinfo{journal}{Nature (London)} \textbf{\bibinfo{volume}{454}},
  \bibinfo{pages}{744} (\bibinfo{year}{2008}).

\bibitem[{\citenamefont{{Jin} et~al.}(2009)\citenamefont{{Jin}, {Stewart}, and
  {Gaebler}}}]{jin2009}
\bibinfo{author}{\bibfnamefont{D.~S.} \bibnamefont{{Jin}}},
  \bibinfo{author}{\bibfnamefont{J.~T.} \bibnamefont{{Stewart}}},
  \bibnamefont{and} \bibinfo{author}{\bibfnamefont{J.~P.}
  \bibnamefont{{Gaebler}}}, in \emph{\bibinfo{booktitle}{Pushing the Frontiers
  of Atomic Physics}}, edited by
  \bibinfo{editor}{\bibnamefont{{R.~C{\^o}t{\'e}, P.~L.~Gould, M.~Rozman, \&
  W.~W.~Smith}}} (\bibinfo{publisher}{World Scientific Publishing Co.},
  \bibinfo{address}{Singapore}, \bibinfo{year}{2009}), pp.
  \bibinfo{pages}{213--219}.

\bibitem[{\citenamefont{Meden and Sch{\"o}nhammer}(1992)}]{meden_spectral}
\bibinfo{author}{\bibfnamefont{V.}~\bibnamefont{Meden}} \bibnamefont{and}
  \bibinfo{author}{\bibfnamefont{K.}~\bibnamefont{Sch{\"o}nhammer}},
  \bibinfo{journal}{Phys. Rev. B} \textbf{\bibinfo{volume}{46}},
  \bibinfo{pages}{15753} (\bibinfo{year}{1992}).

\bibitem[{\citenamefont{Voit}(1993)}]{voit_spectral}
\bibinfo{author}{\bibfnamefont{J.}~\bibnamefont{Voit}}, \bibinfo{journal}{Phys.
  Rev. B} \textbf{\bibinfo{volume}{47}}, \bibinfo{pages}{6740}
  (\bibinfo{year}{1993}).

\bibitem[{\citenamefont{Nakamura and Suzumura}(1997)}]{nakamura_suzumura}
\bibinfo{author}{\bibfnamefont{N.}~\bibnamefont{Nakamura}} \bibnamefont{and}
  \bibinfo{author}{\bibfnamefont{Y.}~\bibnamefont{Suzumura}},
  \bibinfo{journal}{Prog. Theor. Phys.} \textbf{\bibinfo{volume}{98}},
  \bibinfo{pages}{29} (\bibinfo{year}{1997}).

\bibitem[{\citenamefont{Penc et~al.}(1996)\citenamefont{Penc, Hallberg, Mila,
  and Shiba}}]{penc_shadowband}
\bibinfo{author}{\bibfnamefont{K.}~\bibnamefont{Penc}},
  \bibinfo{author}{\bibfnamefont{K.}~\bibnamefont{Hallberg}},
  \bibinfo{author}{\bibfnamefont{F.}~\bibnamefont{Mila}}, \bibnamefont{and}
  \bibinfo{author}{\bibfnamefont{H.}~\bibnamefont{Shiba}},
  \bibinfo{journal}{Phys. Rev. Lett.} \textbf{\bibinfo{volume}{77}},
  \bibinfo{pages}{1390} (\bibinfo{year}{1996}).

\bibitem[{\citenamefont{Penc et~al.}(1997)\citenamefont{Penc, Hallberg, Mila,
  and Shiba}}]{penc1997}
\bibinfo{author}{\bibfnamefont{K.}~\bibnamefont{Penc}},
  \bibinfo{author}{\bibfnamefont{K.}~\bibnamefont{Hallberg}},
  \bibinfo{author}{\bibfnamefont{F.}~\bibnamefont{Mila}}, \bibnamefont{and}
  \bibinfo{author}{\bibfnamefont{H.}~\bibnamefont{Shiba}},
  \bibinfo{journal}{Phys. Rev. B} \textbf{\bibinfo{volume}{55}},
  \bibinfo{pages}{15475} (\bibinfo{year}{1997}).

\bibitem[{\citenamefont{Ogata and Shiba}(1990)}]{ogata_inf}
\bibinfo{author}{\bibfnamefont{M.}~\bibnamefont{Ogata}} \bibnamefont{and}
  \bibinfo{author}{\bibfnamefont{H.}~\bibnamefont{Shiba}},
  \bibinfo{journal}{Phys. Rev. B} \textbf{\bibinfo{volume}{41}},
  \bibinfo{pages}{2326} (\bibinfo{year}{1990}).

\bibitem[{\citenamefont{Matveev et~al.}(2007)\citenamefont{Matveev, Furusaki,
  and Glazman}}]{matveev2007}
\bibinfo{author}{\bibfnamefont{K.~A.} \bibnamefont{Matveev}},
  \bibinfo{author}{\bibfnamefont{A.}~\bibnamefont{Furusaki}}, \bibnamefont{and}
  \bibinfo{author}{\bibfnamefont{L.~I.} \bibnamefont{Glazman}},
  \bibinfo{journal}{Phys. Rev. B} \textbf{\bibinfo{volume}{76}},
  \bibinfo{pages}{155440} (\bibinfo{year}{2007}).

\bibitem[{\citenamefont{{Favand} et~al.}(1997)\citenamefont{{Favand}, {Haas},
  {Penc}, {Mila}, and {Dagotto}}}]{favand1997}
\bibinfo{author}{\bibfnamefont{J.}~\bibnamefont{{Favand}}},
  \bibinfo{author}{\bibfnamefont{S.}~\bibnamefont{{Haas}}},
  \bibinfo{author}{\bibfnamefont{K.}~\bibnamefont{{Penc}}},
  \bibinfo{author}{\bibfnamefont{F.}~\bibnamefont{{Mila}}}, \bibnamefont{and}
  \bibinfo{author}{\bibfnamefont{E.}~\bibnamefont{{Dagotto}}},
  \bibinfo{journal}{Phys. Rev. B} \textbf{\bibinfo{volume}{55}},
  \bibinfo{pages}{4859} (\bibinfo{year}{1997}).

\bibitem[{\citenamefont{Zacher et~al.}(1998)\citenamefont{Zacher, Arrigoni,
  Hanke, and Schrieffer}}]{zacher1998}
\bibinfo{author}{\bibfnamefont{M.~G.} \bibnamefont{Zacher}},
  \bibinfo{author}{\bibfnamefont{E.}~\bibnamefont{Arrigoni}},
  \bibinfo{author}{\bibfnamefont{W.}~\bibnamefont{Hanke}}, \bibnamefont{and}
  \bibinfo{author}{\bibfnamefont{J.~R.} \bibnamefont{Schrieffer}},
  \bibinfo{journal}{Phys. Rev. B} \textbf{\bibinfo{volume}{57}},
  \bibinfo{pages}{6370} (\bibinfo{year}{1998}).

\bibitem[{\citenamefont{Abendschein and Assaad}(2006)}]{abendschein2006}
\bibinfo{author}{\bibfnamefont{A.}~\bibnamefont{Abendschein}} \bibnamefont{and}
  \bibinfo{author}{\bibfnamefont{F.~F.} \bibnamefont{Assaad}},
  \bibinfo{journal}{Phys. Rev. B} \textbf{\bibinfo{volume}{73}},
  \bibinfo{pages}{165119} (\bibinfo{year}{2006}).

\bibitem[{\citenamefont{Benthien et~al.}(2004)\citenamefont{Benthien, Gebhard,
  and Jeckelmann}}]{benthien2004}
\bibinfo{author}{\bibfnamefont{H.}~\bibnamefont{Benthien}},
  \bibinfo{author}{\bibfnamefont{F.}~\bibnamefont{Gebhard}}, \bibnamefont{and}
  \bibinfo{author}{\bibfnamefont{E.}~\bibnamefont{Jeckelmann}},
  \bibinfo{journal}{Phys. Rev. Lett.} \textbf{\bibinfo{volume}{92}},
  \bibinfo{pages}{256401} (\bibinfo{year}{2004}).

\bibitem[{\citenamefont{Bulut et~al.}(2006)\citenamefont{Bulut, Matsueda,
  Tohyama, and Maekawa}}]{bulut2006}
\bibinfo{author}{\bibfnamefont{N.}~\bibnamefont{Bulut}},
  \bibinfo{author}{\bibfnamefont{H.}~\bibnamefont{Matsueda}},
  \bibinfo{author}{\bibfnamefont{T.}~\bibnamefont{Tohyama}}, \bibnamefont{and}
  \bibinfo{author}{\bibfnamefont{S.}~\bibnamefont{Maekawa}},
  \bibinfo{journal}{Phys. Rev. B} \textbf{\bibinfo{volume}{74}},
  \bibinfo{pages}{113106} (\bibinfo{year}{2006}).

\bibitem[{\citenamefont{Miyashita et~al.}(2002)\citenamefont{Miyashita,
  Kawaguchi, and Kawakami}}]{Miyashita2002}
\bibinfo{author}{\bibfnamefont{S.}~\bibnamefont{Miyashita}},
  \bibinfo{author}{\bibfnamefont{A.}~\bibnamefont{Kawaguchi}},
  \bibnamefont{and} \bibinfo{author}{\bibfnamefont{N.}~\bibnamefont{Kawakami}},
  \bibinfo{journal}{J. Phys. Soc. Jpn.} \textbf{\bibinfo{volume}{71}},
  \bibinfo{pages}{1947} (\bibinfo{year}{2002}).

\bibitem[{\citenamefont{{Rabello} and {Si}}(2002)}]{Rabello2002}
\bibinfo{author}{\bibfnamefont{S.}~\bibnamefont{{Rabello}}} \bibnamefont{and}
  \bibinfo{author}{\bibfnamefont{Q.}~\bibnamefont{{Si}}},
  \bibinfo{journal}{Europhys. Lett.} \textbf{\bibinfo{volume}{60}},
  \bibinfo{pages}{882} (\bibinfo{year}{2002}).

\bibitem[{\citenamefont{Feiguin and Huse}(2009)}]{feiguin2009}
\bibinfo{author}{\bibfnamefont{A.~E.} \bibnamefont{Feiguin}} \bibnamefont{and}
  \bibinfo{author}{\bibfnamefont{D.~A.} \bibnamefont{Huse}},
  \bibinfo{journal}{Phys. Rev. B} \textbf{\bibinfo{volume}{79}},
  \bibinfo{pages}{100507} (\bibinfo{year}{2009}).

\bibitem[{\citenamefont{Muttalib and Emery}(1986)}]{muttalib1986}
\bibinfo{author}{\bibfnamefont{K.~A.} \bibnamefont{Muttalib}} \bibnamefont{and}
  \bibinfo{author}{\bibfnamefont{V.~J.} \bibnamefont{Emery}},
  \bibinfo{journal}{Phys. Rev. Lett.} \textbf{\bibinfo{volume}{57}},
  \bibinfo{pages}{1370} (\bibinfo{year}{1986}).

\bibitem[{\citenamefont{Frahm and Korepin}(1991)}]{frahm_confinv_field}
\bibinfo{author}{\bibfnamefont{H.}~\bibnamefont{Frahm}} \bibnamefont{and}
  \bibinfo{author}{\bibfnamefont{V.~E.} \bibnamefont{Korepin}},
  \bibinfo{journal}{Phys. Rev. B} \textbf{\bibinfo{volume}{43}},
  \bibinfo{pages}{5653} (\bibinfo{year}{1991}).

\bibitem[{\citenamefont{Penc and S{\'o}lyom}(1993)}]{penc_magnetic_field}
\bibinfo{author}{\bibfnamefont{K.}~\bibnamefont{Penc}} \bibnamefont{and}
  \bibinfo{author}{\bibfnamefont{J.}~\bibnamefont{S{\'o}lyom}},
  \bibinfo{journal}{Phys. Rev. B} \textbf{\bibinfo{volume}{47}},
  \bibinfo{pages}{6273} (\bibinfo{year}{1993}).

\bibitem[{\citenamefont{{Orso}}(2007)}]{orso2007}
\bibinfo{author}{\bibfnamefont{G.}~\bibnamefont{{Orso}}},
  \bibinfo{journal}{Phys. Rev. Lett.} \textbf{\bibinfo{volume}{98}},
  \bibinfo{pages}{070402} (\bibinfo{year}{2007}).

\bibitem[{\citenamefont{Cazalilla and Ho}(2003)}]{cazalilla03_mixture}
\bibinfo{author}{\bibfnamefont{M.~A.} \bibnamefont{Cazalilla}}
  \bibnamefont{and} \bibinfo{author}{\bibfnamefont{A.~F.} \bibnamefont{Ho}},
  \bibinfo{journal}{Phys. Rev. Lett.} \textbf{\bibinfo{volume}{91}},
  \bibinfo{pages}{150403} (\bibinfo{year}{2003}).

\bibitem[{\citenamefont{{Mathey} et~al.}(2004)\citenamefont{{Mathey}, {Wang},
  {Hofstetter}, {Lukin}, and {Demler}}}]{mathey2004}
\bibinfo{author}{\bibfnamefont{L.}~\bibnamefont{{Mathey}}},
  \bibinfo{author}{\bibfnamefont{D.-W.} \bibnamefont{{Wang}}},
  \bibinfo{author}{\bibfnamefont{W.}~\bibnamefont{{Hofstetter}}},
  \bibinfo{author}{\bibfnamefont{M.~D.} \bibnamefont{{Lukin}}},
  \bibnamefont{and} \bibinfo{author}{\bibfnamefont{E.}~\bibnamefont{{Demler}}},
  \bibinfo{journal}{Physical Review Letters} \textbf{\bibinfo{volume}{93}},
  \bibinfo{pages}{120404} (\bibinfo{year}{2004}).

\bibitem[{\citenamefont{Gogolin et~al.}(1999)\citenamefont{Gogolin, Nersesyan,
  and Tsvelik}}]{gogolin_1dbook}
\bibinfo{author}{\bibfnamefont{A.~O.} \bibnamefont{Gogolin}},
  \bibinfo{author}{\bibfnamefont{A.~A.} \bibnamefont{Nersesyan}},
  \bibnamefont{and} \bibinfo{author}{\bibfnamefont{A.~M.}
  \bibnamefont{Tsvelik}}, \emph{\bibinfo{title}{Bosonization and Strongly
  Correlated Systems}} (\bibinfo{publisher}{Cambridge University Press},
  \bibinfo{address}{Cambridge}, \bibinfo{year}{1999}).

\bibitem[{\citenamefont{Orignac et~al.}(2010)\citenamefont{Orignac, Tsuchiizu,
  and Suzumura}}]{orignac2010_mix}
\bibinfo{author}{\bibfnamefont{E.}~\bibnamefont{Orignac}},
  \bibinfo{author}{\bibfnamefont{M.}~\bibnamefont{Tsuchiizu}},
  \bibnamefont{and} \bibinfo{author}{\bibfnamefont{Y.}~\bibnamefont{Suzumura}},
  \bibinfo{journal}{Phys. Rev. A} \textbf{\bibinfo{volume}{81}},
  \bibinfo{pages}{053626} (\bibinfo{year}{2010}).

\bibitem[{\citenamefont{Hikihara et~al.}(2005)\citenamefont{Hikihara, Furusaki,
  and Matveev}}]{hikihara2005}
\bibinfo{author}{\bibfnamefont{T.}~\bibnamefont{Hikihara}},
  \bibinfo{author}{\bibfnamefont{A.}~\bibnamefont{Furusaki}}, \bibnamefont{and}
  \bibinfo{author}{\bibfnamefont{K.~A.} \bibnamefont{Matveev}},
  \bibinfo{journal}{Phys. Rev. B} \textbf{\bibinfo{volume}{72}},
  \bibinfo{pages}{035301} (\bibinfo{year}{2005}).

\bibitem[{\citenamefont{{Le Bellac}}(1992)}]{lebellac_qft}
\bibinfo{author}{\bibfnamefont{M.}~\bibnamefont{{Le Bellac}}},
  \emph{\bibinfo{title}{Quantum and Statistical Field Theory}}
  (\bibinfo{publisher}{Oxford University Press}, \bibinfo{address}{Oxford, UK},
  \bibinfo{year}{1992}).

\bibitem[{\citenamefont{Haldane}(1981)}]{haldane_bosonisation}
\bibinfo{author}{\bibfnamefont{F.~D.~M.} \bibnamefont{Haldane}},
  \bibinfo{journal}{J. Phys. C} \textbf{\bibinfo{volume}{14}},
  \bibinfo{pages}{2585} (\bibinfo{year}{1981}).

\bibitem[{\citenamefont{Abramowitz and
  Stegun}(1972)}]{abramowitz_math_functions}
\bibinfo{author}{\bibfnamefont{M.}~\bibnamefont{Abramowitz}} \bibnamefont{and}
  \bibinfo{author}{\bibfnamefont{I.}~\bibnamefont{Stegun}},
  \emph{\bibinfo{title}{Handbook of mathematical functions}}
  (\bibinfo{publisher}{Dover}, \bibinfo{address}{New York},
  \bibinfo{year}{1972}).

\bibitem[{\citenamefont{{Erd\'elyi} et~al.}(1953)\citenamefont{{Erd\'elyi},
  Magnus, Oberhettinger, and Tricomi}}]{erdelyi_functions_1}
\bibinfo{author}{\bibfnamefont{A.}~\bibnamefont{{Erd\'elyi}}},
  \bibinfo{author}{\bibfnamefont{W.}~\bibnamefont{Magnus}},
  \bibinfo{author}{\bibfnamefont{F.}~\bibnamefont{Oberhettinger}},
  \bibnamefont{and} \bibinfo{author}{\bibfnamefont{F.~G.}
  \bibnamefont{Tricomi}}, \emph{\bibinfo{title}{Higher transcendental
  functions}}, vol.~\bibinfo{volume}{1} (\bibinfo{publisher}{McGraw-Hill},
  \bibinfo{address}{NY}, \bibinfo{year}{1953}).

\bibitem[{\citenamefont{{Olde Daalhuis} and Askey}(2010)}]{dlmf_appell}
\bibinfo{author}{\bibfnamefont{A.~B.} \bibnamefont{{Olde Daalhuis}}}
  \bibnamefont{and} \bibinfo{author}{\bibfnamefont{R.~A.} \bibnamefont{Askey}},
  \emph{\bibinfo{title}{Digital library of mathematical functions, ch. 16}},
  \bibinfo{howpublished}{http://dlmf.nist.gov/16.15} (\bibinfo{year}{2010}),
  \bibinfo{note}{{National Institute of Standards and Technology,
  \textit{Release date 2010-05-07}}}.

\bibitem[{\citenamefont{{Iucci} et~al.}(2007)\citenamefont{{Iucci}, {Fiete},
  and {Giamarchi}}}]{iucci2007}
\bibinfo{author}{\bibfnamefont{A.}~\bibnamefont{{Iucci}}},
  \bibinfo{author}{\bibfnamefont{G.~A.} \bibnamefont{{Fiete}}},
  \bibnamefont{and}
  \bibinfo{author}{\bibfnamefont{T.}~\bibnamefont{{Giamarchi}}},
  \bibinfo{journal}{Phys. Rev. B} \textbf{\bibinfo{volume}{75}},
  \bibinfo{pages}{205116} (\bibinfo{year}{2007}).

\bibitem[{\citenamefont{Gaudin}(1967)}]{gaudin_fermions}
\bibinfo{author}{\bibfnamefont{M.}~\bibnamefont{Gaudin}},
  \bibinfo{journal}{Phys. Lett. A} \textbf{\bibinfo{volume}{24}},
  \bibinfo{pages}{55} (\bibinfo{year}{1967}).

\bibitem[{\citenamefont{Yang}(1967)}]{yang_fermions}
\bibinfo{author}{\bibfnamefont{C.~N.} \bibnamefont{Yang}},
  \bibinfo{journal}{Phys. Rev. Lett.} \textbf{\bibinfo{volume}{19}},
  \bibinfo{pages}{1312} (\bibinfo{year}{1967}).

\bibitem[{\citenamefont{Kawakami and Yang}(1990)}]{kawakami_tj}
\bibinfo{author}{\bibfnamefont{N.}~\bibnamefont{Kawakami}} \bibnamefont{and}
  \bibinfo{author}{\bibfnamefont{S.~K.} \bibnamefont{Yang}},
  \bibinfo{journal}{Phys. Rev. Lett.} \textbf{\bibinfo{volume}{65}},
  \bibinfo{pages}{2309} (\bibinfo{year}{1990}).

\bibitem[{\citenamefont{{Imambekov} and {Demler}}(2006)}]{imambekov2006}
\bibinfo{author}{\bibfnamefont{A.}~\bibnamefont{{Imambekov}}} \bibnamefont{and}
  \bibinfo{author}{\bibfnamefont{E.}~\bibnamefont{{Demler}}},
  \bibinfo{journal}{Ann. Phys. (N. Y.)} \textbf{\bibinfo{volume}{321}},
  \bibinfo{pages}{2390} (\bibinfo{year}{2006}).

\bibitem[{\citenamefont{{Luscher} and {Laeuchli}}(2009)}]{luscher2009}
\bibinfo{author}{\bibfnamefont{A.}~\bibnamefont{{Luscher}}} \bibnamefont{and}
  \bibinfo{author}{\bibfnamefont{A.}~\bibnamefont{{Laeuchli}}},
  \emph{\bibinfo{title}{{Imbalanced thee-component Fermi gas with attractive
  interactions: Multiple FFLO-pairing, Bose-Fermi and Fermi-Fermi mixtures
  versus collapse and phase separation}}},
  \bibinfo{howpublished}{arXiv:0906.0768} (\bibinfo{year}{2009}).

\bibitem[{\citenamefont{Gradshteyn and Ryzhik}(1980)}]{gradshteyn80_tables}
\bibinfo{author}{\bibfnamefont{A.}~\bibnamefont{Gradshteyn}} \bibnamefont{and}
  \bibinfo{author}{\bibfnamefont{R.}~\bibnamefont{Ryzhik}},
  \emph{\bibinfo{title}{Tables of integrals series and products}}
  (\bibinfo{publisher}{Academic Press}, \bibinfo{address}{New-York},
  \bibinfo{year}{1980}).

\bibitem[{\citenamefont{{Di Francesco} et~al.}(1997)\citenamefont{{Di
  Francesco}, Mathieu, and Senechal}}]{difrancesco_book_conformal}
\bibinfo{author}{\bibfnamefont{P.}~\bibnamefont{{Di Francesco}}},
  \bibinfo{author}{\bibfnamefont{P.}~\bibnamefont{Mathieu}}, \bibnamefont{and}
  \bibinfo{author}{\bibfnamefont{D.}~\bibnamefont{Senechal}},
  \emph{\bibinfo{title}{Conformal Field Theory}}
  (\bibinfo{publisher}{Springer-Verlag}, \bibinfo{address}{Berlin},
  \bibinfo{year}{1997}).

\bibitem[{\citenamefont{Cardy}(1996)}]{cardy_conformal_book}
\bibinfo{author}{\bibfnamefont{J.}~\bibnamefont{Cardy}},
  \emph{\bibinfo{title}{Scaling and Renormalization in Statistical Physics}}
  (\bibinfo{publisher}{Cambridge University Press},
  \bibinfo{address}{Cambridge}, \bibinfo{year}{1996}).

\bibitem[{\citenamefont{Saleur}(1998)}]{saleur_houches}
\bibinfo{author}{\bibfnamefont{H.}~\bibnamefont{Saleur}}, in
  \emph{\bibinfo{booktitle}{Topological aspects of low dimensional systems}},
  edited by \bibinfo{editor}{\bibfnamefont{A.}~\bibnamefont{Comtet}},
  \bibinfo{editor}{\bibfnamefont{T.}~\bibnamefont{Jolicoeur}},
  \bibinfo{editor}{\bibfnamefont{S.}~\bibnamefont{Ouvry}}, \bibnamefont{and}
  \bibinfo{editor}{\bibfnamefont{F.}~\bibnamefont{David}}
  (\bibinfo{publisher}{Springer}, \bibinfo{address}{Berlin},
  \bibinfo{year}{1998}), vol.~\bibinfo{volume}{69} of
  \emph{\bibinfo{series}{Les Houches Summer School}}, p. \bibinfo{pages}{475}.

\bibitem[{\citenamefont{Frahm and Korepin}(1990)}]{frahm_confinv}
\bibinfo{author}{\bibfnamefont{H.}~\bibnamefont{Frahm}} \bibnamefont{and}
  \bibinfo{author}{\bibfnamefont{V.~E.} \bibnamefont{Korepin}},
  \bibinfo{journal}{Phys. Rev. B} \textbf{\bibinfo{volume}{42}},
  \bibinfo{pages}{10553} (\bibinfo{year}{1990}).

\bibitem[{\citenamefont{Senechal}(2003)}]{senechal_bosonization_revue}
\bibinfo{author}{\bibfnamefont{D.}~\bibnamefont{Senechal}}, in
  \emph{\bibinfo{booktitle}{Theoretical Methods for Strongly Correlated
  Electrons}}, edited by
  \bibinfo{editor}{\bibfnamefont{D.}~\bibnamefont{{S{\'e}nechal {\it et al.}}}}
  (\bibinfo{publisher}{Springer}, \bibinfo{address}{New York},
  \bibinfo{year}{2003}), CRM Series in Mathematical Physics,
  \bibinfo{note}{cond-mat/9908262}.

\end{thebibliography}
\end{document}